\documentclass[twocolumn]{aastex631}
\usepackage{comment}
\usepackage{amsmath}
\usepackage[mathlines]{lineno}

\usepackage{graphicx}
\usepackage{float}
\usepackage{soul}
\begin{document}
\title{Quantifying Suppression of Solar Surface Magnetic Flux Advection with Increasing Field Strength}
\author[0000-0002-0824-3109]{V.\ Aparna}
\affiliation{Lockheed-Martin Solar \& Astrophysics Laboratory, 3251 Hanover Street, Building 203, Palo Alto, CA 94306, USA}
\affiliation{Bay Area Environmental Research Institute, NASA Research Park, Moffett Field, CA 94035, USA}
\author[0000-0001-7817-2978]{Sanjiv K.\ Tiwari}
\affiliation{Lockheed-Martin Solar \& Astrophysics Laboratory, 3251 Hanover Street, Building 203, Palo Alto, CA 94306, USA}
\affiliation{Bay Area Environmental Research Institute, NASA Research Park, Moffett Field, CA 94035, USA}
\author[0000-0002-5691-6152]{Ronald L.\ Moore}
\affiliation{NASA Marshall Space Flight Center, Huntsville, AL 35812, USA}
\author[0000-0001-7620-362X]{Navdeep K.\ Panesar}
\affiliation{Lockheed-Martin Solar \& Astrophysics Laboratory, 3251 Hanover Street, Building 203, Palo Alto, CA 94306, USA}
\affiliation{Bay Area Environmental Research Institute, NASA Research Park, Moffett Field, CA 94035, USA}
\author[0000-0003-2244-641X]{Brian Welsch}
\affiliation{Natural \& Applied Sciences, University of Wisconsin-Green Bay, 2420 Nicolet Drive, Green Bay, WI 54311, USA}
\author[0000-0002-8370-952X]{Bart De Pontieu}
\affiliation{Lockheed-Martin Solar \& Astrophysics Laboratory, 3251 Hanover Street, Building 203, Palo Alto, CA 94306, USA}
\author[0000-0003-2622-7310]{Aimee Norton}
\affiliation{HEPL Solar Physics, Stanford University, Stanford, CA, 94305-4085, USA}

\begin{abstract}

One of the main theories for heating of the solar corona is based on the idea that solar convection shuffles and tangles magnetic field lines to make many small-scale current sheets that, via reconnection, heat coronal loops.  
\citet{2017ApJ...843L..20T} present evidence that, besides depending on loop length and other factors, the brightness of a coronal loop depends on the field strength in the loop’s feet and the freedom of convection in the feet.  
While it is known that strong solar magnetic fields suppress convection, the decrease in the speed of horizontal advection of magnetic flux with increasing field strength has not been quantified before. We quantify that trend by analyzing 24-hours of HMI-SHARP vector magnetograms of each of six sunspot active regions and their surroundings. Using Fourier Local Correlation Tracking, we estimate the horizontal advection speed of the magnetic flux at each pixel in which the vertical component of the magnetic field strength (Bz) is well above ($\ge$ 150 G) noise level. 
We find that the average horizontal advection speed of magnetic flux steadily decreases as Bz increases, from $110 \pm 3$ m/s for 150 G (in network and plage) to $10 \pm 4$ m/s for 2500 G (in sunspot umbra). The trend is well fit by a fourth degree polynomial. These results quantitatively confirm the expectation that magnetic flux advection is suppressed by increasing magnetic field strength. The presented quantitative relation should be useful for future MHD simulations of coronal heating. 
\end{abstract}

\section{Introduction} \label{sec:intro}

Convection in the Sun consists of hot plasma rising from the solar interior to the surface, cooling as it ascends, and sinking back to the bottom of the convection zone. This process not only sustains the Sun's luminosity, but is essential to the solar dynamo \citep{2009LRSP....6....2N,2012LRSP....9....4S}. 
In the presence of magnetic field, convective motions are inhibited by the Lorentz force, particularly in the direction perpendicular to the magnetic field \citep{2008MNRAS.387..698B}. This suppression is strong in sunspot umbrae, %\textcolor{purple}{where granulation cells are absent} 
where the magnetic pressure greatly exceeds the gas pressure \citep{2011LRSP....8....3R}. Around the umbrae, in the penumbrae, convection cells are subject to the relatively weaker fields there that produce elongated filamentary structures, known as penumbral filaments (e.g., \citealt{2011ApJ...729....5R,2013A&A...557A..25T}, and references therein).  

Convective flows on the time-scales of granular, meso and super-granular motions sweep the flux in the quiet regions towards the inter-granular and super-granular network lanes \citep{1988ApJ...327..964S,1989ApJ...336..475T,2012LRSP....9....4S}.  
In the quiet Sun, fields are easily moved around by the convective flows. 
How much the magnetic field increasingly slows the convection with increasing field-strength, from weak (quiet Sun) to strong (sunspots), is yet uncertain, requiring detailed observations and modeling efforts. 
 
The suppression of convection by magnetic field has been studied in the context of sunspots, plages, and quiet regions individually and using data of different cadences and types, to measure horizontal and vertical flow speeds. However, there has been no study that has uniformly measured how much horizontal advection of magnetic flux is suppressed as field strength increases across the entire range of field strength over the solar surface. 

\citet{1989ApJ...336..475T} performed local correlation tracking analysis on a plage region surrounded by quiet regions using continuum images from the Swedish Solar Telescope (SST). They found the mean horizontal speeds in granulation to be $\sim$370 m/s in ``field-free" interior of network cells. \citet{1992ApJ...393..782T} measured the speed as a function of line-of-sight field strength. The speed decreased from 275 to 100 m/s from weak to strong fields. In the overlapping range of field strength of their data and ours, the trend they found is consistent with the trend we find, shown later. 
\citet{2000SoPh..193..313S} also performed LCT over a quiet supergranular field of view using MDI continuum images. The mean advection speed from images taken over 45 hours with 1-minute cadence averaged hourly 
was found to be about 488 m/s, and the maximum speed about 1.5 km/s. \citet{2020A&A...633A..67K} estimated a diverging flow speed of about $\sim$0.8 - 1km/s in a quiet region patch on the Sun with emerging flux with G-band images separated by $\sim$62 min. The velocities derived from correlation tracking depend on the spatial resolution and cadence of the measurements.  
Most studies that have used LCT to study granular and supergranular flows to obtain flow speeds, to study the evolution, or the role of sub-surface and surface flows in the advection of magnetic elements, have used white-light or G-band images of the photosphere \citep{2000SoPh..192..351D,2000SoPh..193..299H,2002A&A...387..672K,2018LRSP...15....6R,2022ApJ...926..127M} with the measured speeds in quiet regions ranging between 300-500 m/s. 

Thus, flow rates on the Sun vary depending on the location, magnetic field strength, and the observed feature \citep{2024ApJ...973...11P}, with higher speeds if the observed region is active and producing flares \citep{2019RSPTA.37780387A,2020ApJ...897L..23K}. 
Furthermore, flows having large ranges of spatial scale and duration are present on the Sun, which challenge measuring the full range of speeds. Various tracking algorithms based on Local Correlation Tracking (LCT; \citealt{1988ApJ...333..427N}) are commonly used for tracking features on the Sun and measuring their flow speeds. The results of these measurements depend on the time interval between consecutive images and on the spatial width of the kernel used for correlations of consecutive images to measure the proper-motion shifts of the features. Due to the combined effects of all these factors, it is difficult to directly compare the flow speeds between measurements of different types of regions using different types of data. In this work we establish a general relation that applies to magneto-convection across all types of regions using similar data. 

The work reported here quantifies the increasing suppression of horizontal advection of solar surface magnetic flux with increasing strength of the vertical component of the magnetic field. We assume that the horizontal advection of the flux is driven by horizontal convective plasma flows. The speed we measure is the horizontal advection speed of clumps of magnetic flux that are wider than several granules.
In our future work, we will use the empirical curve (of the measured decreasing horizontal advection speed as a function of increasing field strength) obtained here to investigate the heating of coronal magnetic loops in solar active regions. There, we will use the curve obtained here as a proxy for the increasing suppression of horizontal advection of magnetic flux with increasing vertical field strength at granular and intergranular scales.

Here, we apply Fourier Local Correlation Tracking (FLCT; \citealt{2008ASPC..383..373F}) to magnetograms of six non-flaring active regions (ARs) to quantify the increasing suppression of horizontal advection of magnetic flux with increasing vertical field strength by measuring the horizontal speed of magnetic flux elements. These measurements are from datasets that are uniform in spatial resolution and cadence, allowing direct comparison of the horizontal speeds. We describe the AR data and the LCT method used in the next section. We present our results in section \ref{sec:results}, followed by a discussion of the results in section \ref{sec:disc}.

  \begin{figure*}%[H] %[!htbp]
    \centering
    \includegraphics[trim={8cm 0.5cm 4cm 0},clip,scale=0.16]{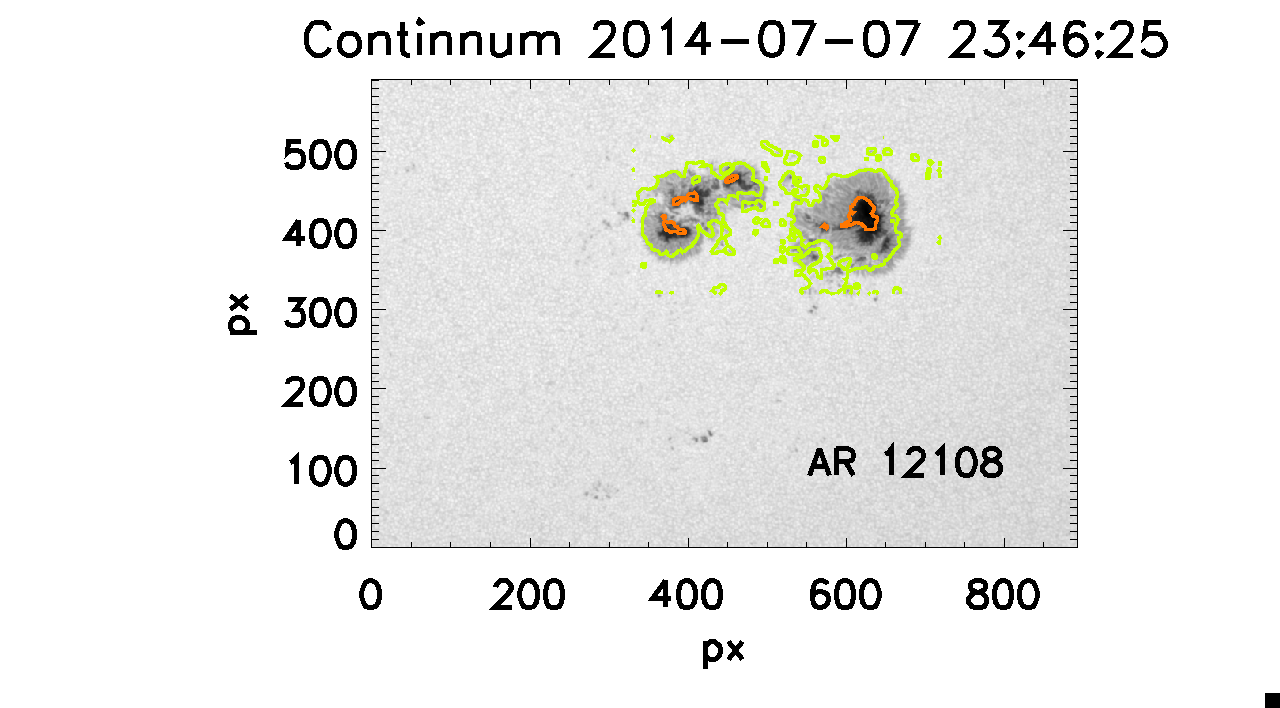}
    \includegraphics[trim={3.5cm 0.5cm 3cm 0},clip,scale=0.16]{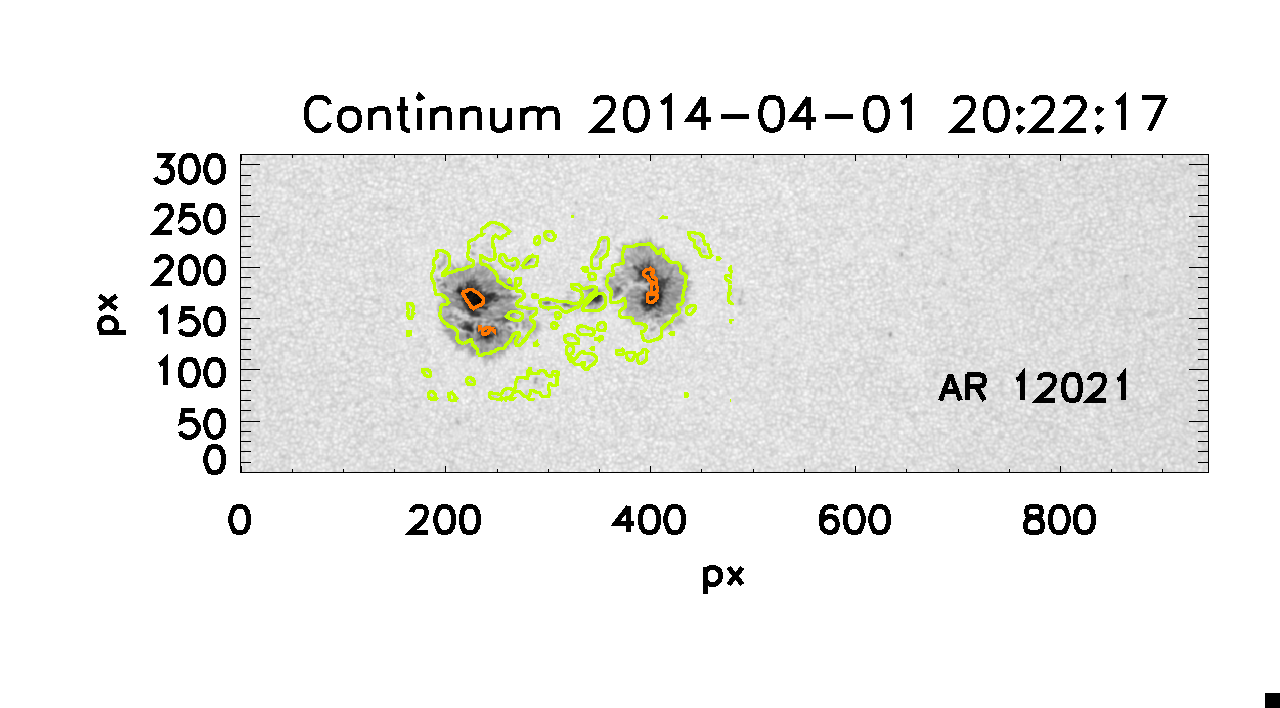}
    \includegraphics[trim={6cm 0.5cm 3.6cm 0},clip,scale=0.16]{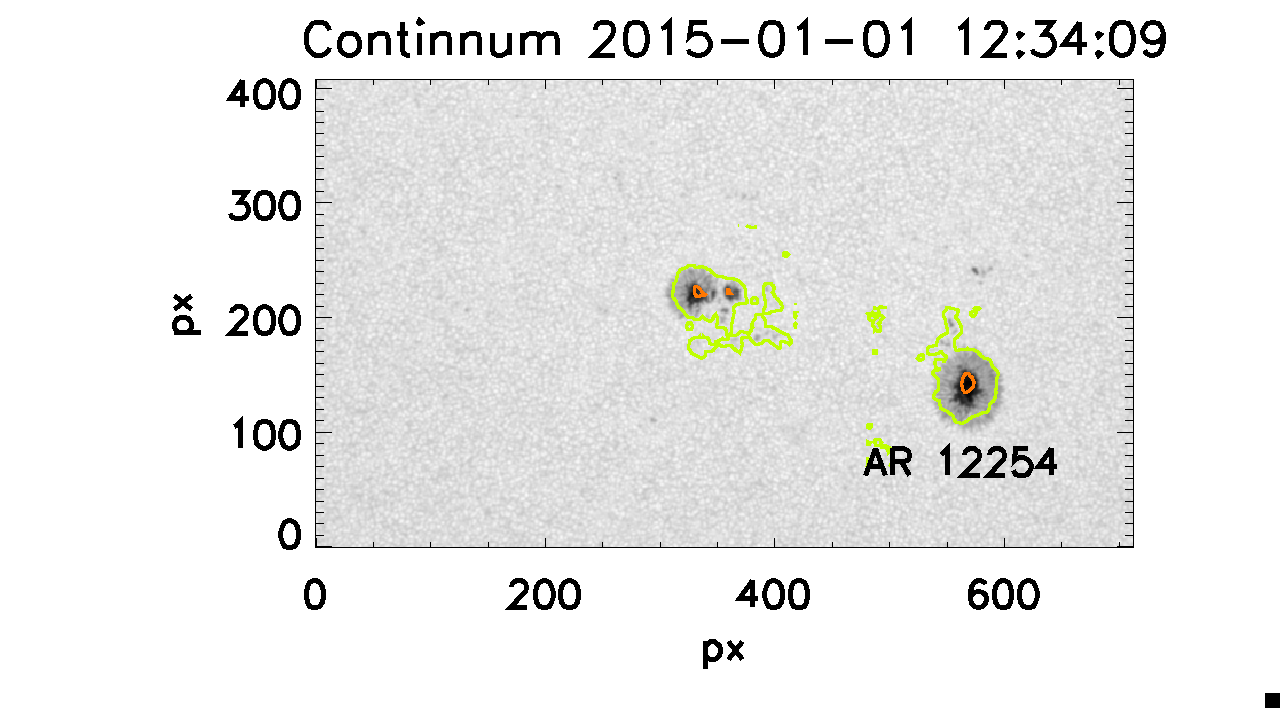}
    \includegraphics[trim={8cm 0.5cm 4cm 0},clip,scale=0.16]{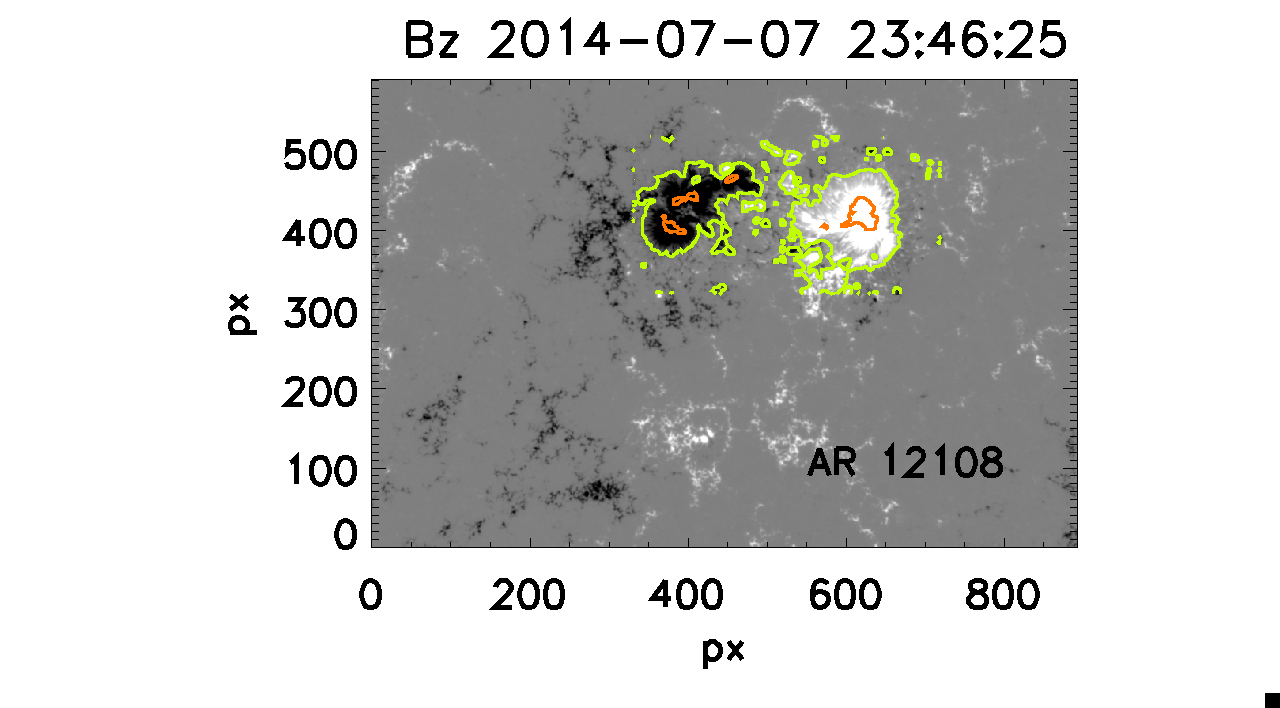}
    \includegraphics[trim={3.5cm 0.5cm 3cm 0},clip,scale=0.16]{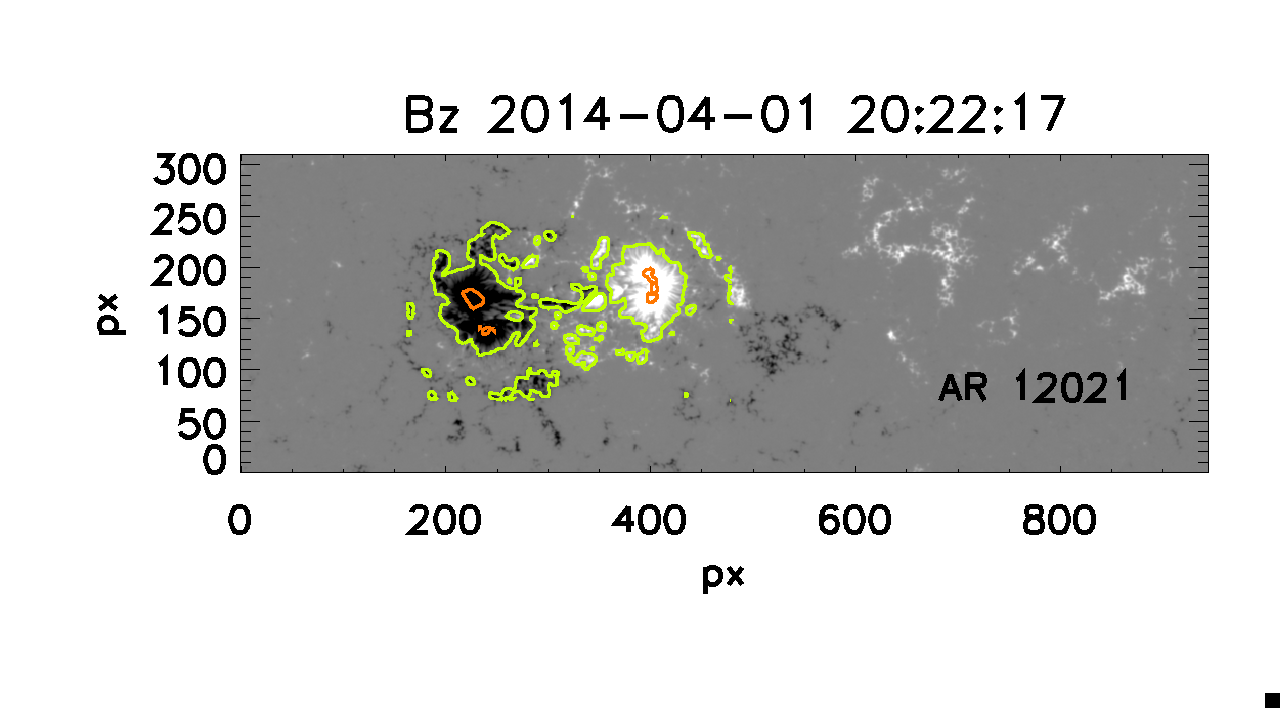}
    \includegraphics[trim={6cm 0.5cm 5cm 0},clip,scale=0.16]{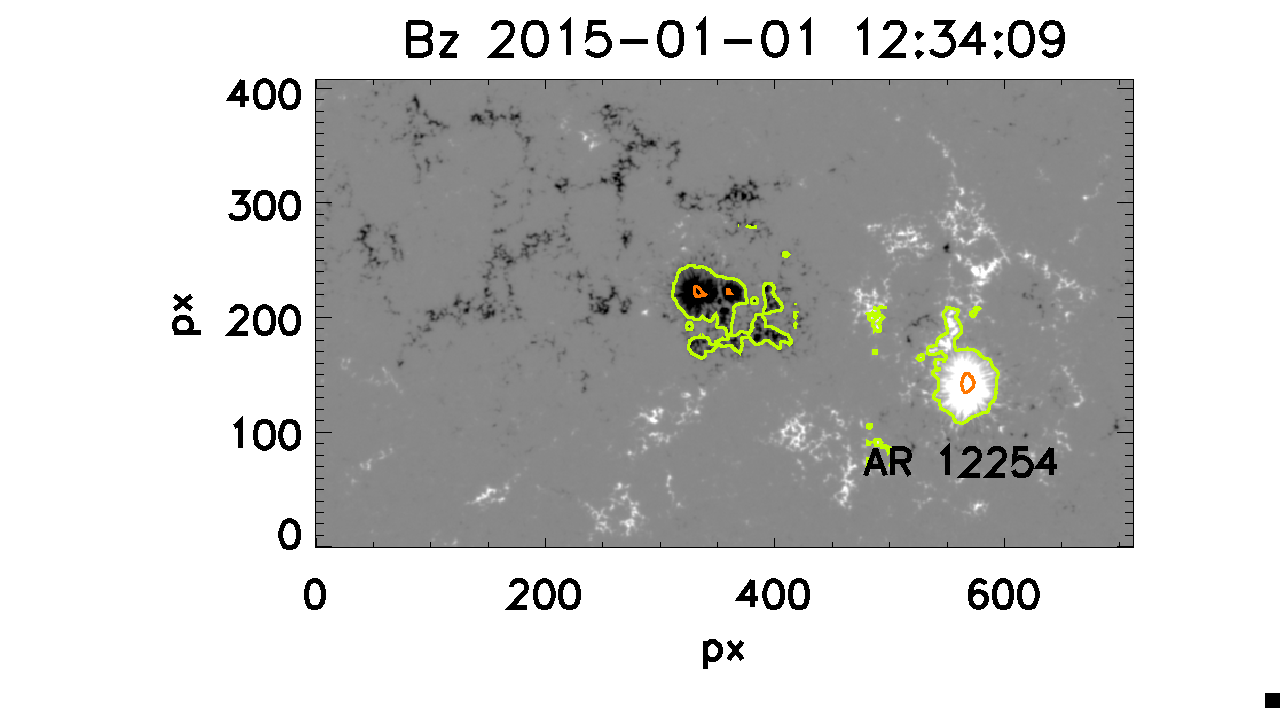}

    \includegraphics[trim={7cm 0.5cm 4cm 0},clip,scale=0.155]{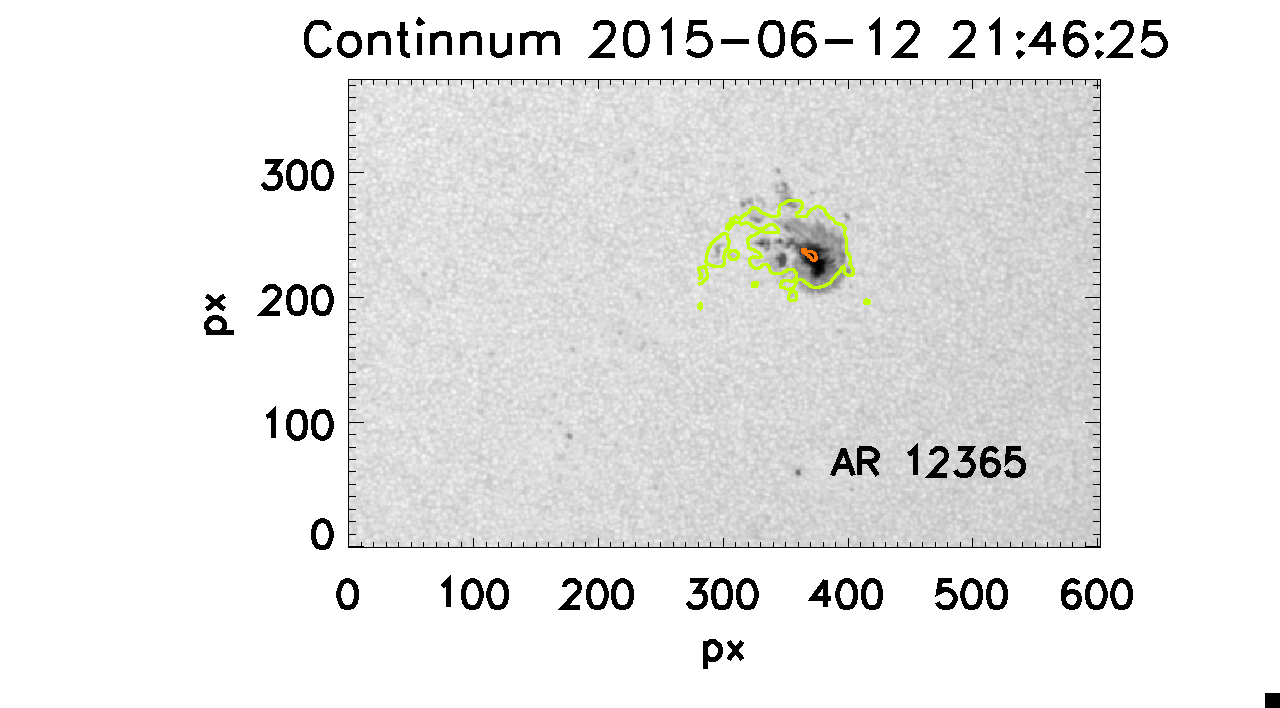}
    \includegraphics[trim={3.5cm 0.5cm 2.5cm 0},clip,scale=0.155]{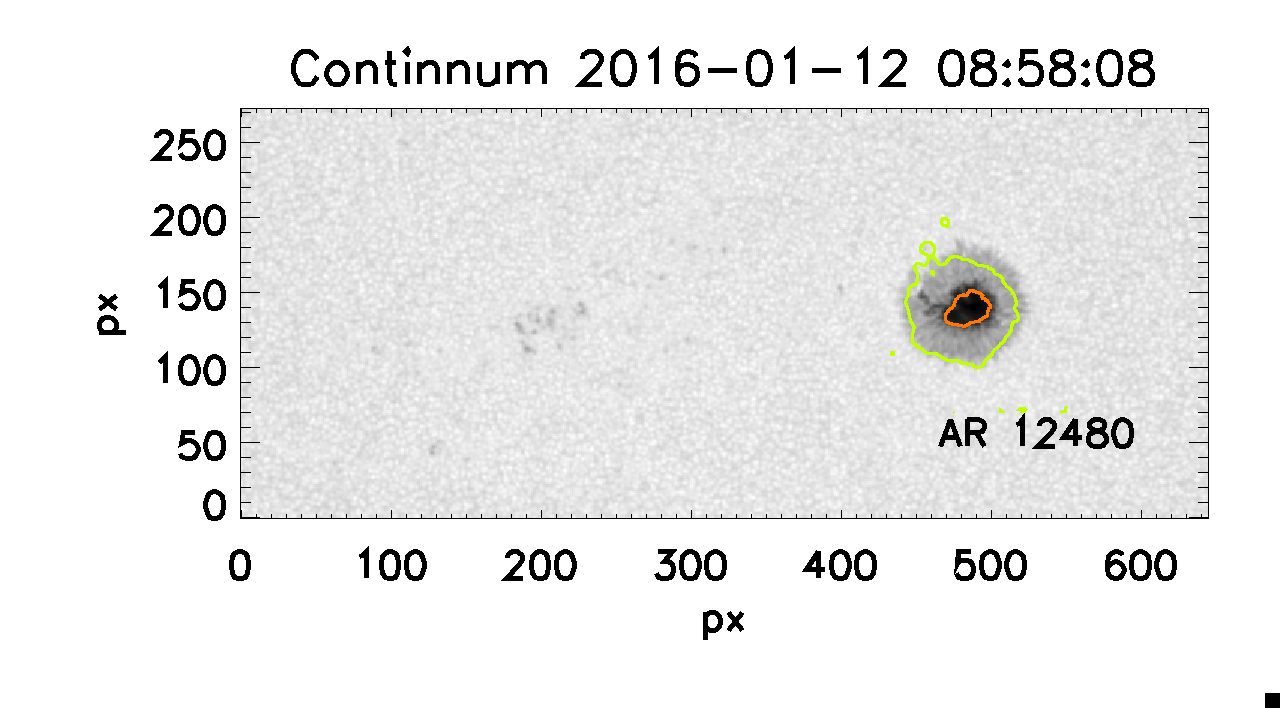}
    \includegraphics[trim={3.5cm 0.5cm 1.5cm 0},clip,scale=0.155]{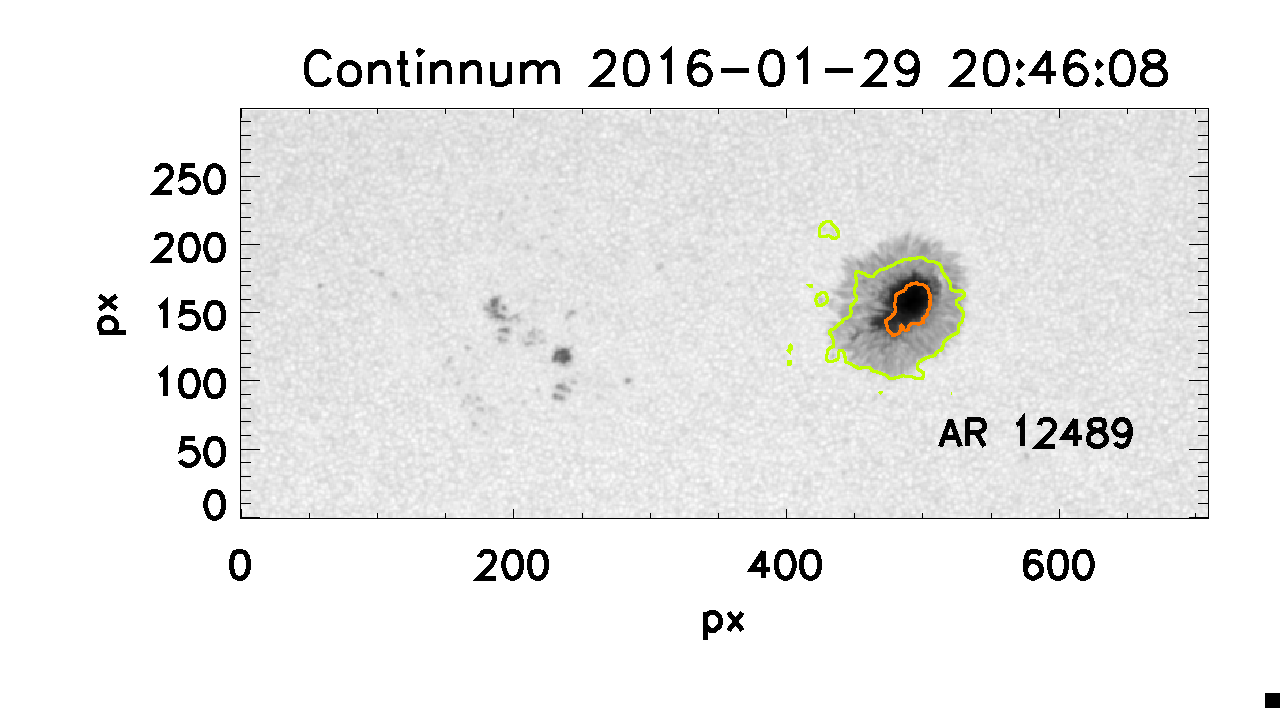}
    \includegraphics[trim={7cm 0.5cm 5cm 0},clip,scale=0.155]{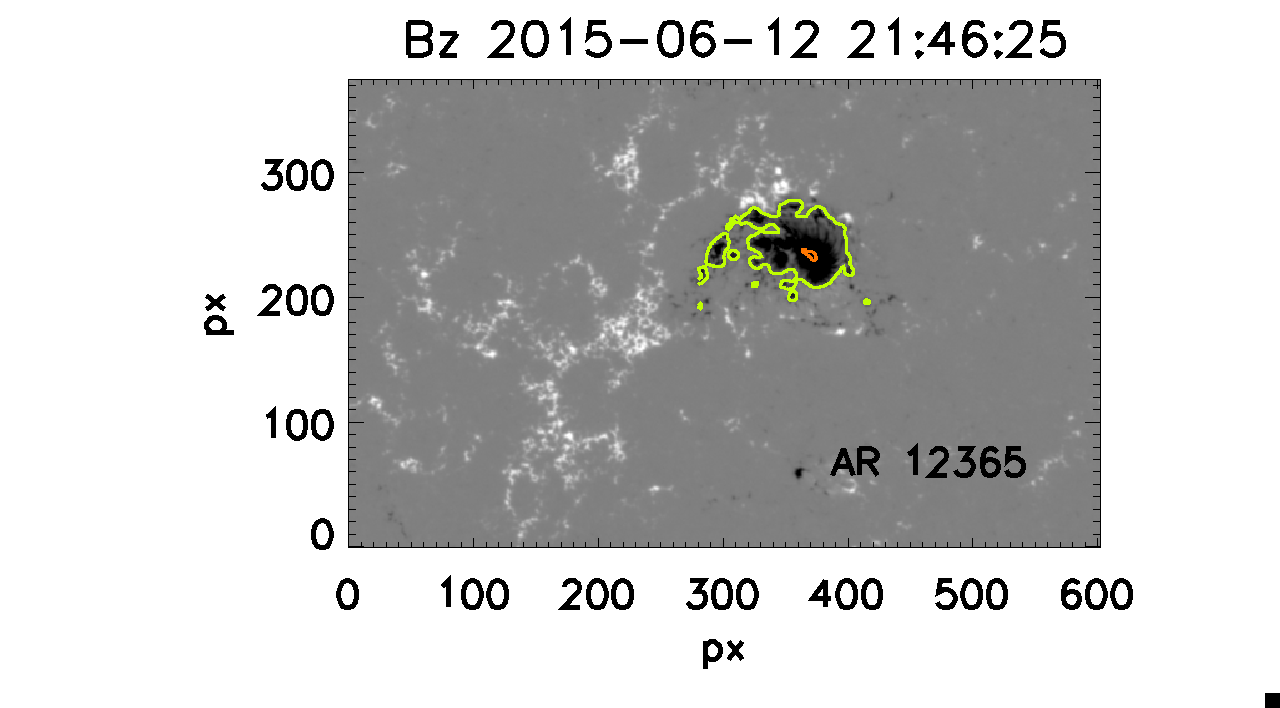}
    \includegraphics[trim={3.5cm 0.5cm 2.5cm 0},clip,scale=0.155]{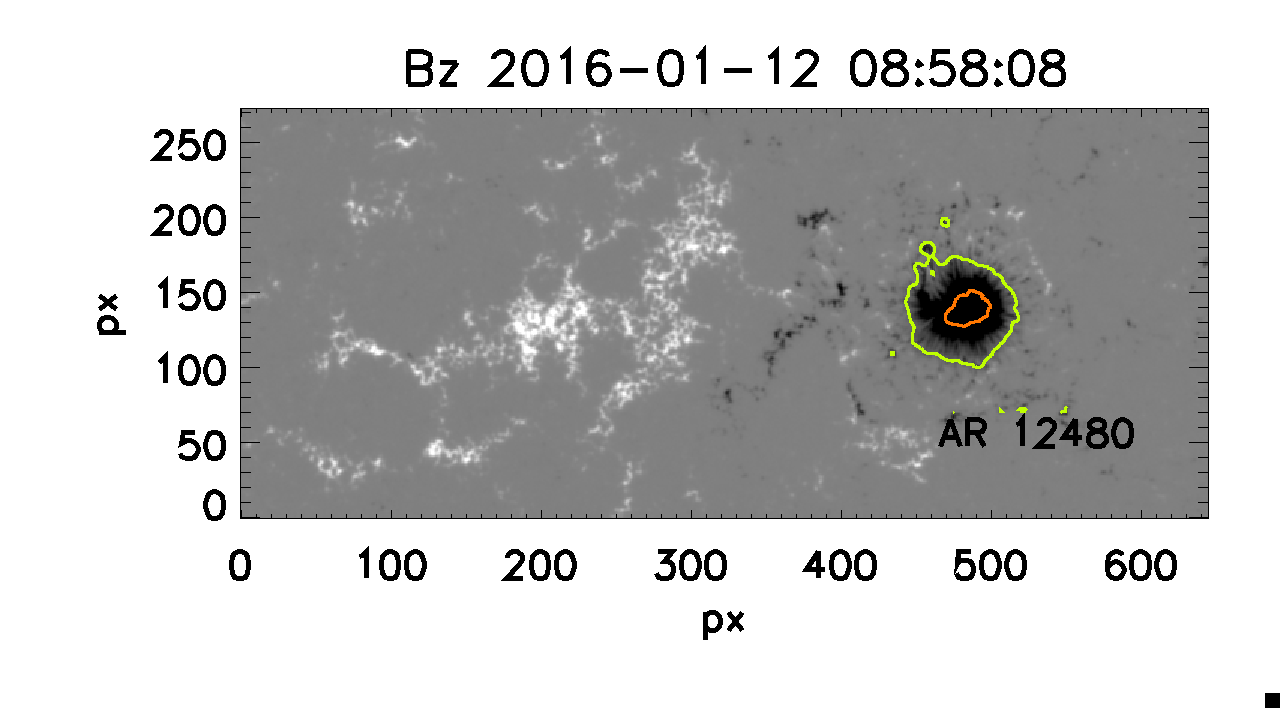}
    \includegraphics[trim={3.5cm 0.5cm 1.5cm 0},clip,scale=0.155]{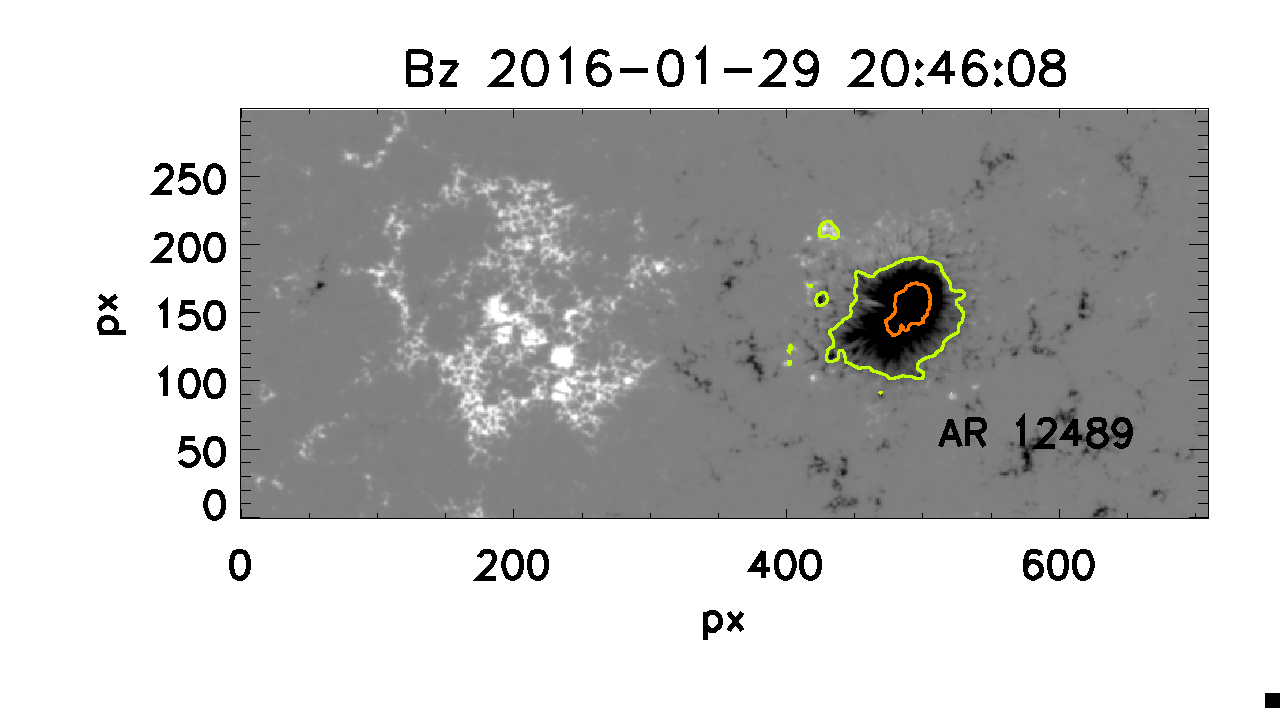}    
    \caption{Two pairs of rows, each displaying continuum intensity images and Bz magnetograms (vertical magnetic field) from SDO/HMI, for our six active regions are shown. The top pair of rows shows our three SS (sunspot-sunspot) type ARs and the bottom pair of rows shows our three SP (sunspot-plage) type ARs. Overlaid on each image are contours of magnetic field strength (Bz) roughly tracing the boundaries between the penumbra and umbra (of strength 1500G depicted in orange) and the outer penumbra (of strength 200G depicted in fluorescent yellow-green). They are drawn such that the umbra-penumbra Bz contours are well within the umbra to ensure that it outlines the stronger umbral field strengths and the outer-penumbral Bz contours are near but inside the edge of the sunspot. Note that the plage and network regions away from the sunspots also have many Bz values of 200 G but are not marked.}   
    \label{fig:SP_contContours}
\end{figure*}

\begin{figure*}%[H] %[!htbp]
%\begin{interactive}{animation}{animations/2014Jul07_Bz.mp4}
%figure call (e.g.\plotone, \includegraphics, etc.)
\includegraphics[trim={1.7cm 9cm 1cm 2cm},clip,scale=0.92]{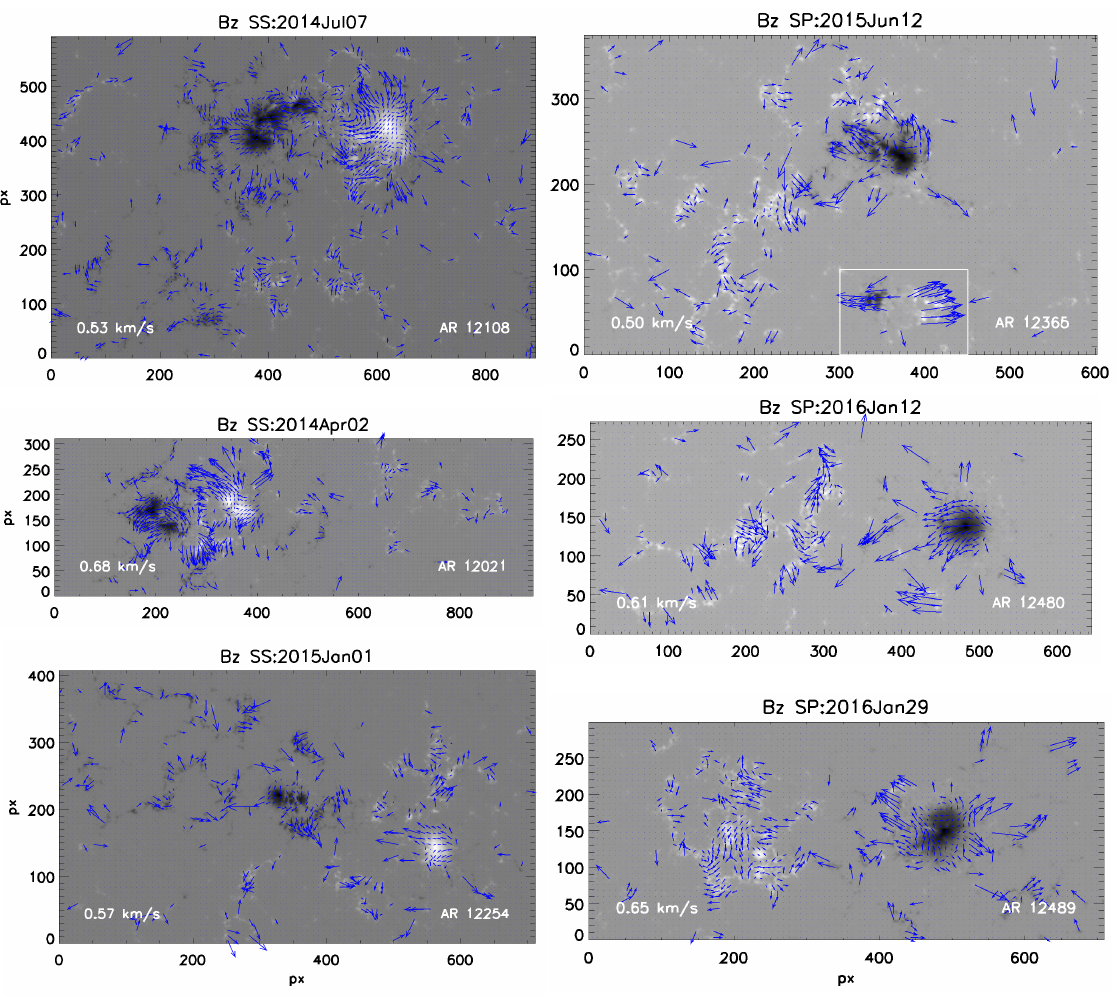}
%\end{interactive}
 \caption{Horizontal advection velocity arrows overlaid on a Bz magnetogram at one sample during the 24 hours for each AR. The panels on the left show the plots for SS type ARs and those on the right show the plots for the SP type ARs. The speed of the longest arrow in each frame is given in the legend at the bottom left in each panel. A close-up view of the area marked in the white box in the top right panel is given in the Appendix. A movie for each AR is available online. Each movie spans 24-hours and has a duration of 12 seconds and begins 12 hours before and ends 12 hours after the central meridian time (given in Table \ref{tab:ARtype}). The animations show the movement of photospheric magnetic fluxes in time and the overlaid arrows point in the direction of this motion in every frame. }
\label{fig:arrowsPlot}
\end{figure*}

\section{Data and Methods}\label{sec:data}

\subsection{Data}

We primarily analyze magnetogram data from the Helioseismic Magnetic Imager (HMI, \citealt{2012SoPh..275..207S}) onboard the Solar Dynamics Observatory (SDO, \citealt{2012SoPh..275....3P}), processed to obtain CEA (Cylindrical Equal Area projection) Bz magnetograms from SHARP (Space-weather HMI Active Region Patch, \citealt{2014SoPh..289.3549B}) vector magnetograms. SHARP magnetograms have a cadence of 12-minutes. Line-of-sight magnetograms (BLOS), available at 45-second intervals, are used for comparing results with those obtained using Bz. We selected six non-flaring ARs for this study. Both datasets for each AR are 24 hours long. 
The SHARP dataset are active region patches with observables including vector magnetic field data that have been disambiguated using the Minimum Energy Code \citep{2009SoPh..260...83L} to resolve the 180$^\circ$ ambiguity in the magnetic field azimuthal component. 
We use a spatial scale of 360 km/pixel in the CEA magnetograms (please refer to \citealt{2012SoPh..275..285C,2013arXiv1309.2392S,2013SoPh..287..279L} for the details of processing the raw polarimetric data to obtain CEA magnetograms). 
The BLOS magnetograms are not deprojected to the disk center. 
In addition to 45-second, we sample them at 3-minute, 6-minute and 12-minute intervals. They are differentially derotated to the time closest to the central meridian crossing of the AR. 
The FLCT code is applied to all six active regions for tracking the magnetic elements and obtaining their horizontal speeds. The active regions (ARs) used for the study are listed in Table \ref{tab:ARtype} and are selected such that their near real-time heliocentric positions are within 30$^{\circ}$ of the solar disk-center and they do not produce major flares for the 24 hours during our observations and have not produced a flare stronger than a C1-class 24 hours before and after the duration of the selected time frames. This is to avoid any lateral displacements in magnetic features caused by large flares (e.g., \citealt{2009ApJ...706L.240G}). As listed in Table \ref{tab:ARtype}, the selected ARs have not produced a flare larger than a C1-class flare during our observations. The selected ARs are also stable during the time of our observations without any large changes in the flux due to emergence or decay.

The ARs used in this study have at least one sunspot each and they are categorized as Sunspot-Sunspot (SS) or Sunspot-Plage (SP) ARs. We have three ARs of each category. Figure \ref{fig:SP_contContours} shows the six ARs in continuum and magnetogram images. 
In the first category, both leading and the following polarities have sunspots. In the second category, the leading polarity contains a sunspot and the trailing polarity contains only plage. 
The magnetograms are also selected such that the middle of the observations coincides with the central meridian passage of the AR. This avoids large projection effects (e.g., \citealt{2016ApJ...833L..31F}). 
Including sunspots, plage regions, and nearby quiet regions provides a wide range of field strengths over which convective motions can be quantified. The speed of advection of the magnetic flux by surface flows is measured from magnetograms using FLCT, with details of the methodology given in the next subsection. 
To check that the speeds obtained using the CEA Bz dataset are reliable, we validate them against the speeds obtained from the 12-minute BLOS dataset. Additionally, because the dataset used by \citealt{1992ApJ...393..782T} had a cadence of $\sim$60-seconds, we perform FLCT with the 45-seconds HMI BLOS dataset in order to compare our results with the \citet{1992ApJ...393..782T} results. For experimental purposes and completeness, FLCT is performed with 3 and 6-minute datasets as well. 
We note that \citealt{1992ApJ...393..782T} performed LCT on ground-based observations of white light images that had a higher spatial resolution than that of HMI magnetograms. 
The results from our analysis of the Bz data are presented in the next section, and those from BLOS are presented in the Appendix. 

\raggedbottom

\begin{table*}%[H] %[!htbp]
\center
    \begin{tabular}{|c|c|c|c|c|c|c|c|}
    \hline
      Index & NOAA AR & Date & AR Type & Position & Max Flare & Central Meridian Time (UTC)\\ 
        \hline
     1 & 12108 & 2014 Jul 07 & SS & S08E02 & -- & 2014 Jul 07 01:00:00 \\
     2 & 12021 & 2014 Apr 02 & SS & S14W05 & C1.1 (Apr 02) & 2014 Apr 01 21:00:00 \\
     3 & 12254 & 2015 Jan 01 & SS & S22W04 & -- & 2015 Jan 01 15:00:00 \\
   % 4 & 12005 & 2014Mar18 & SP & N13W07 & & \\
     4 & 12365 & 2015 Jun 12 & SP & S13W07 & C1.9 & 2015 Jun 11 22:46:00\\         
     5 & 12480 & 2016 Jan 12 & SP & N03W08 & C1.1 & 2016 Jan 12 11:00:00\\
     6 & 12489 & 2016 Jan 29 & SP & N09E01 & C1.0 & 2016 Jan 29 21:30:00 \\
        \hline
    \end{tabular}
    \caption{The largely bipolar active regions used in this study. The table includes the NOAA AR number (second column), date-identifier assigned to each AR in this study (third column), and the AR type (fourth column; SS - AR has opposite polarity sunspots, SP - AR has a leading-polarity sunspot and only plage in its following polarity magnetic flux domain), followed by the AR's location (fifth column) when it was near 0$^\circ$ longitude on the Sun. The sixth column provides the maximum class of flares produced by each AR over its disk passage, except for the second AR, where the flare that occurred during the observation period is given. The last column lists the time when the AR was near near 0$^\circ$ longitude, which serves as the median time for our 24-hour observation window.}
    
 \end{table*} \label{tab:ARtype}

\subsection{Method}

We use %\textcolor{purple}{Fourier Local Correlation Tracking, the method of \cite{2008ASPC..383..373F}} 
FCLT, \citep{2008ASPC..383..373F}, which is similar to the LCT method of \citet{1988ApJ...333..427N}. 
Applying FLCT to magnetograms gives the speed of horizontal advection of the magnetic flux by weighting the region around each pixel using a chosen width of the Gaussian kernel and obtaining the spatial offset in the next frame that gives the best cross-correlation. %\st{with reference to the same pixel in the adjacent frame}.  
The advection speed is estimated by dividing that offset distance %\st{of the pixel position from the cross-correlations and dividing it} 
by the time-interval between the two frames.

Before computing the cross-correlation of the two consecutive weighted magnetograms, FLCT applies a low-pass spatial filter to each of these weighted magnetograms to smooth out the sharp steps across the edges of adjacent pixels. Following \citet{2008ASPC..383..373F}, we use 0.3 for the low-pass filter’s roll-off wave number.

FLCT measures the horizontal speed of flux advection by flows with spatial scales similar to or larger than %that are of the size of 
the FWHM of the Guassian kernel.
We tested the FLCT code on Bz magnetograms with Gaussian widths of 3, 6, 9, 12 and 15 pixels and an FWHM of 15 pixels gave visually the most persistent and least noisy advection speeds. Hence we use 15 pixels FWHM in our measurements.
The length of 15 HMI pixels (7.5$^\prime$$^\prime$) spans $\sim$5.4 Mm on the solar surface, roughly the span of 5 granules. Hence, limited by the CEA Bz dataset's spatial and time resolutions, mesogranular and larger flows are being measured. An example of the measurements of these flows overlaid on a magnetogram of each AR is shown in Fig.\ \ref{fig:arrowsPlot} with the speed of the longest arrow given in the legend in each panel.

We set a threshold value roughly twice the $\sigma$ (estimated uncertainty; \citealt{2012SoPh..279..295L,2014SoPh..289.3483H}) of the magnetogram noise when we apply FLCT to our magnetograms. Our threshold is 150 G for HMI SHARP Bz magnetograms, and 20 G for the BLOS magnetograms. The noise in the Bz magnetograms is about 70 G, much higher than that in the BLOS ones ($\sim7G$) as Bz is derived from processing the Stokes vector products (see \citealt{2014SoPh..289.3549B} and references therein). 
We use Bz magnetograms for our measurements, instead of the $B_{TOTAL}$ (vector-field strength) magnetograms, because (i) by definition, Bz is the magnetic flux through the solar surface per unit area and (ii) the noise in the field’s transverse components gives more noise to B$_{TOTAL}$ than to Bz in regions near disk center such as our six active regions.
FLCT measurements using Bz magnetograms are significantly more persistent than those measured using BLOS or $B_{TOTAL}$ magnetograms (see Appendix), hence we use Bz data for our measurements.

\section{Results}  \label{sec:results}

The horizontal speeds obtained by measuring the horizontal advection of magnetic flux using SHARP Bz data and its decrease with the increasing Bz is the main result of this work and is presented in sections \ref{section3.1} and \ref{sec:FinalPlot}. For each of the six ARs, the horizontal speeds in 50 G-wide bins of increasing Bz are averaged and a mean speed for each bin over a period of 24-hours is obtained. The mean speed in each Bz bin is presented for each AR. 

\subsection{Horizontal Speeds measured using Bz Magnetograms} \label{section3.1}

\begin{figure*}%[H] %[!htbp]
\centering
    \includegraphics[trim={2cm 0.5cm 1cm 0},clip,scale=0.195]{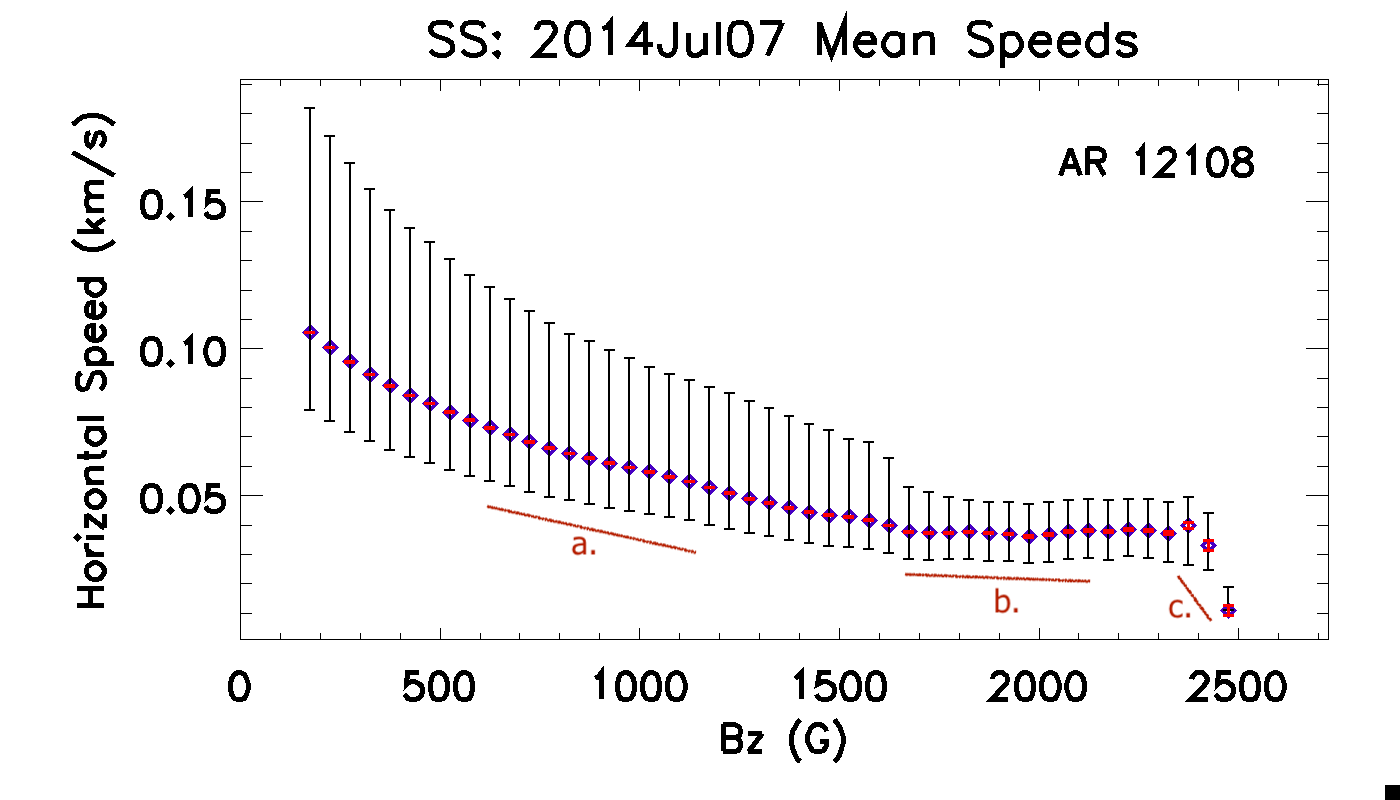}
   \includegraphics[trim={2cm 0.5cm 2cm 0},clip,scale=0.195]{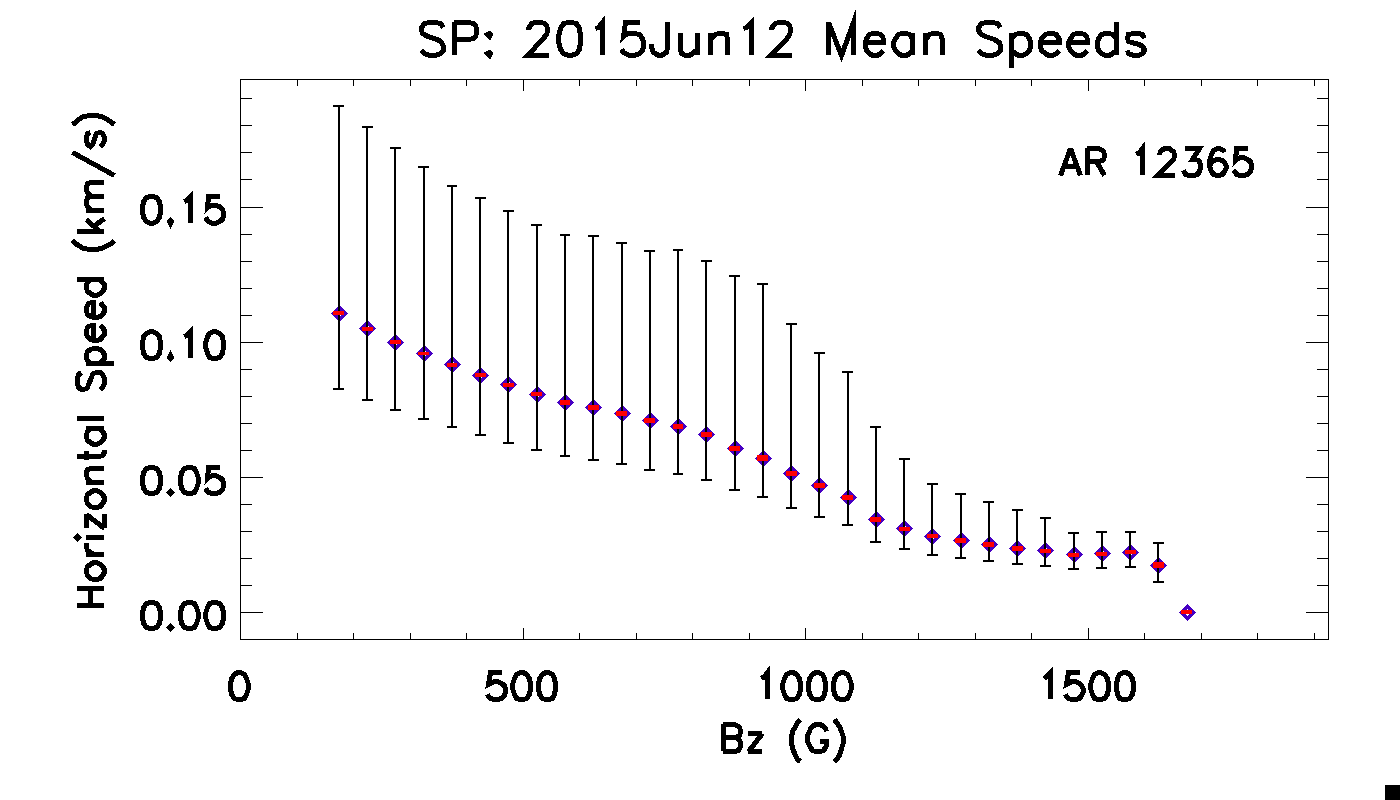}
  
    \includegraphics[trim={2cm 0.5cm 2cm 0},clip,scale=0.195]{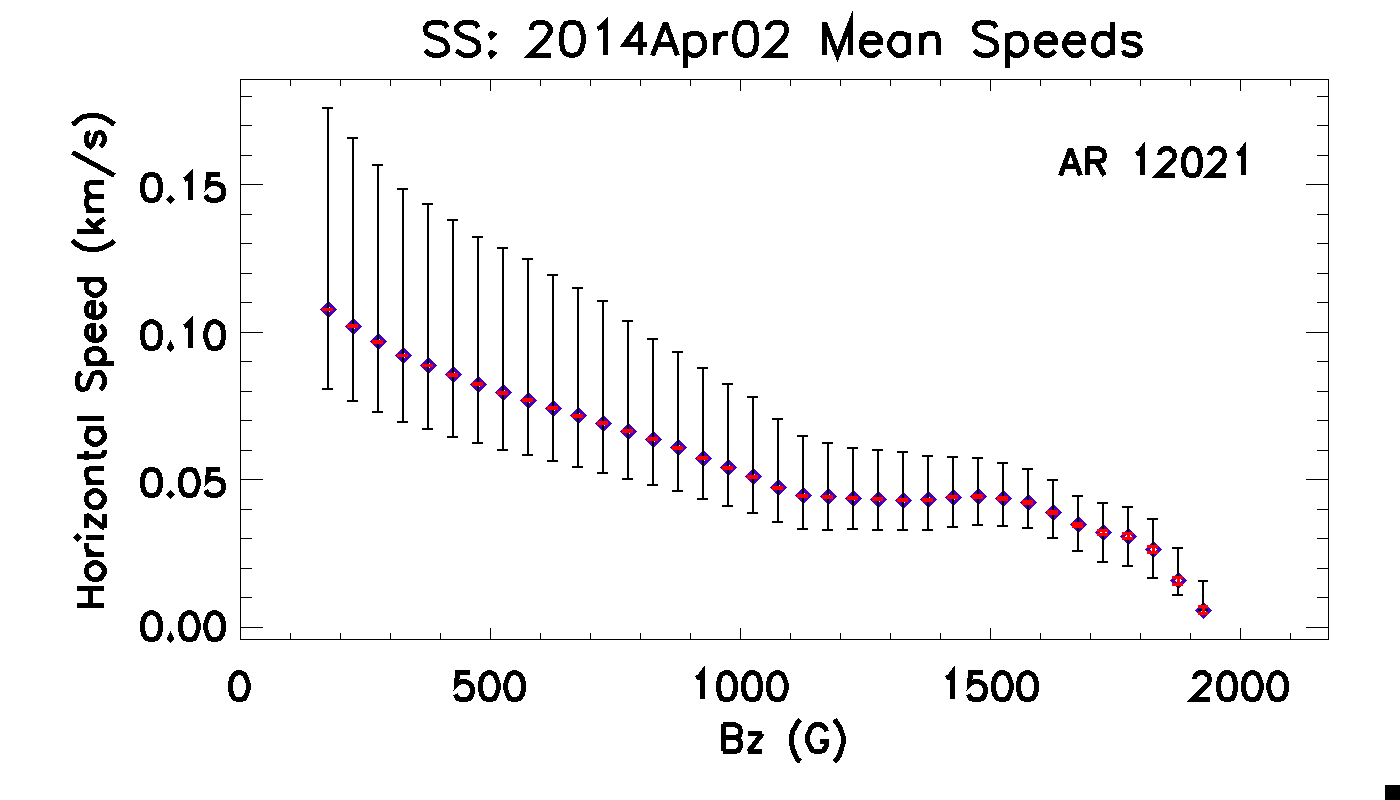}
      \includegraphics[trim={2cm 0.5cm 2cm 0},clip,scale=0.195]{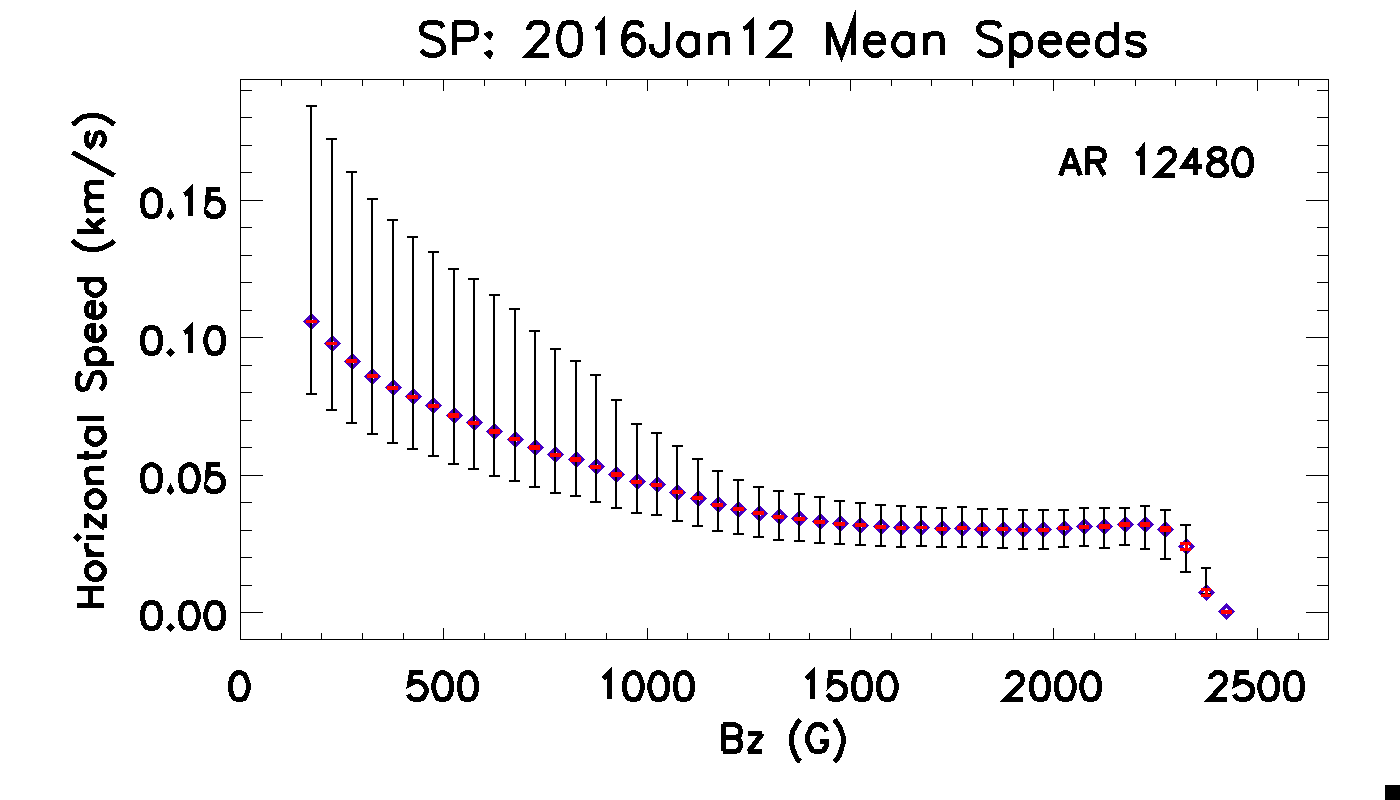}
 
    \includegraphics[trim={2cm 0.5cm 2cm 0},clip,scale=0.195]{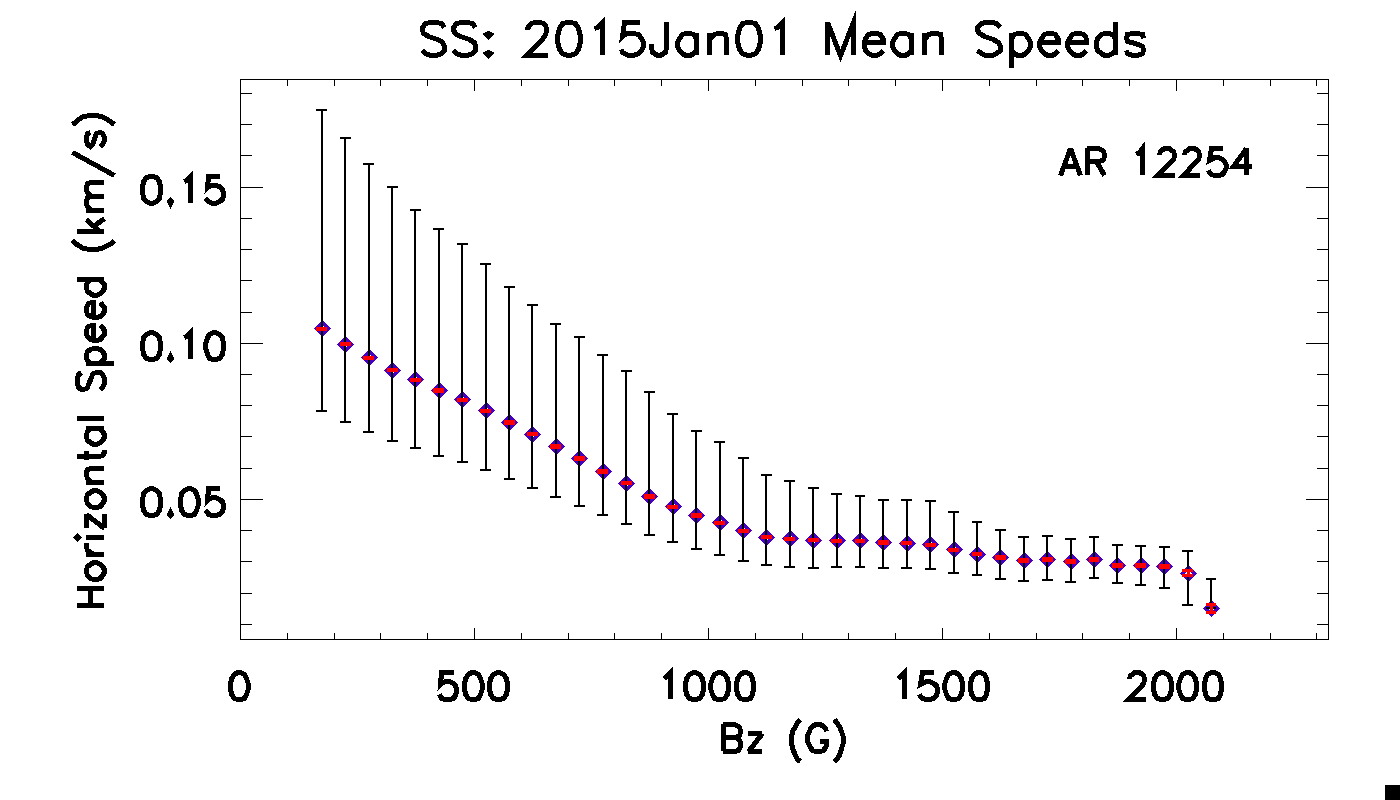}
    \includegraphics[trim={2cm 0.5cm 2cm 0},clip,scale=0.19]{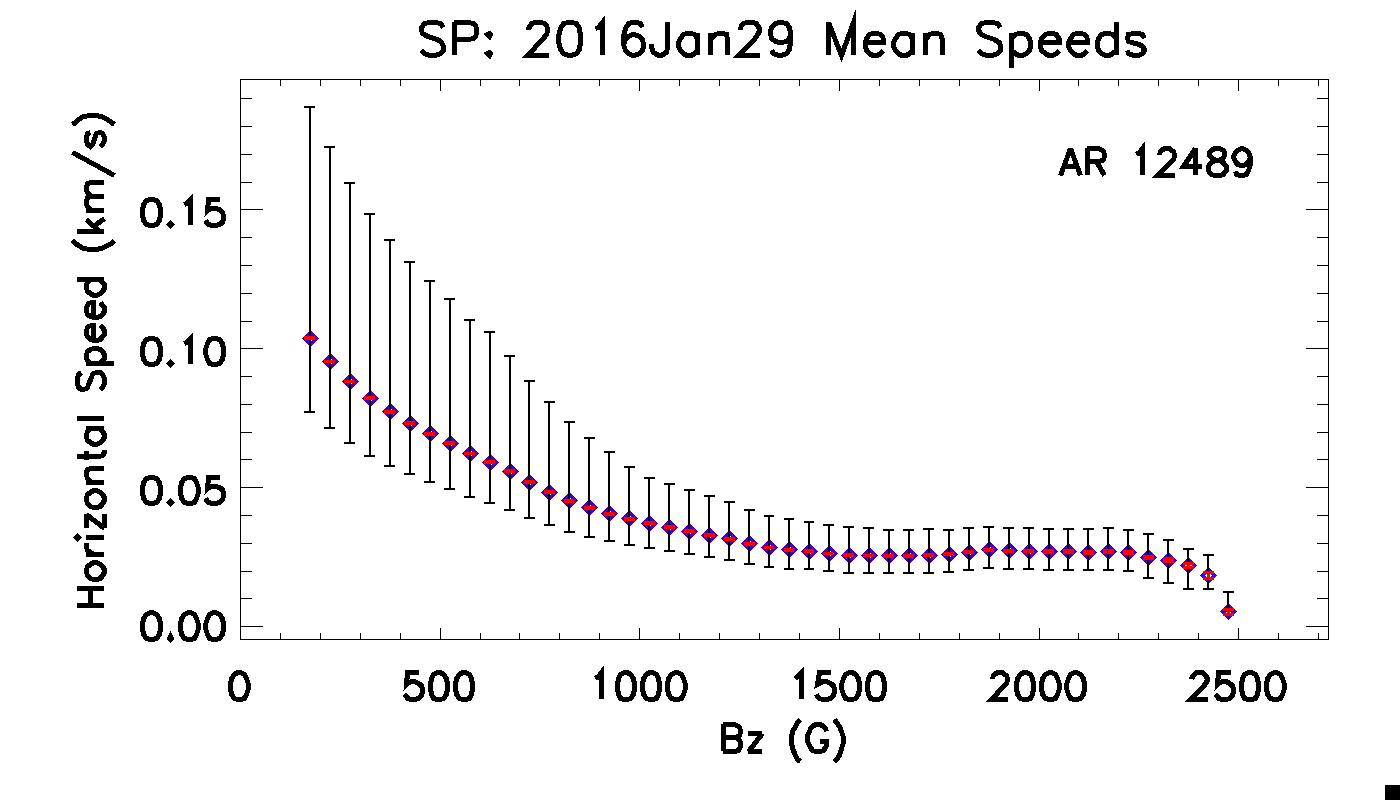}
    \caption{Horizontal speed plotted against Bz for SS (left column) and SP (right column) active regions. The blue diamonds are the mean of the speeds in each 50G-wide bin starting from 150 G. 
    The error bars shown in red show the standard error (the spans of most of the red error bars are within the same span of each plotted diamond and hence show only the error-bar hats). The black error-bars in each panel show the standard deviations of speeds from the mean in each bin, with the upper and lower extents from the mean calculated separately.
   Three maroon line segments named a, b and c in the first panel denote - rapidly decreasing, plateau and the strong-field-end decline in the horizontal speed, which are observed in each AR's plot. }
    \label{fig:horVel} 
\end{figure*}

We present the speed trends from the Bz magnetograms for each of the six ARs in Fig.\ \ref{fig:horVel}. Each point in each panel represents the average of the velocities of all the pixels in the 50G-wide Bz bin, over a duration of 24 hours. Considering the 2$\sigma$ noise level, the horizontal speeds are plotted starting with the 150-200 G bin. Each point is placed at the center of its bin, the first one at 175 G. 
The standard error of each mean and standard deviation from the mean speed are shown with red and black error bars, respectively.
All the six ARs show a strikingly similar general trend with the mean speed decreasing as Bz increases. 

\begin{figure*}%[H] %[!htbp]
        \centering
        \includegraphics[trim={0cm 0cm 0cm 0.0cm},clip,scale=1.0]{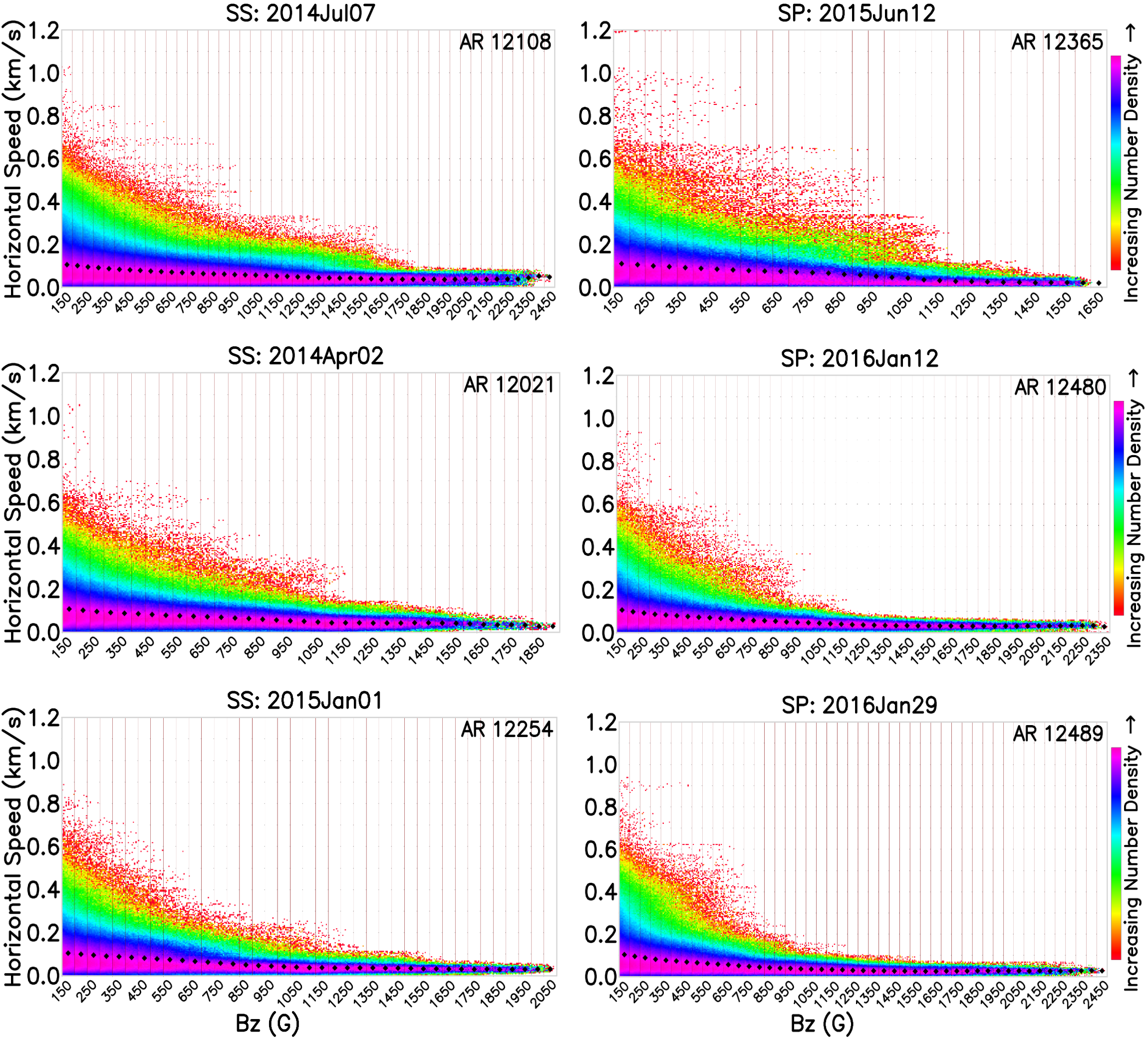} \hfill
        \caption{%Same as Fig.\ \ref{fig:SS_colorsAll} but 
        Each 50G-wide vertical strip in each panel is a 2D histogram of all measured speeds from pixels within the 24-hour period for that 50G bin. The panels in the left column are for the SS ARs, and those in the right column are for the SP ARs. 
        The colorbar on the right of each panel shows that pink is for the highest number density and red is for the lowest. The number density for each color in a bin changes from bin to bin.
        The overlaid black diamonds in each panel is the mean speed curve shown in Fig.\ \ref{fig:horVel}. 
        }
        \label{fig:SS_colorsAll_hist}
\end{figure*}

Similar to the mean speed trends for each AR shown in Fig.\ \ref{fig:horVel}, the overall trend of the speeds for each AR also decreases with increasing field strength. 
Figure \ref{fig:SS_colorsAll_hist} depicts this for the six ARs, wherein a histogram is shown for each of the 50G bins. In all the cases, the number density of measured speeds decreases progressively with increasing speed. 
The number density of measured speeds is greater at slower speeds, near and below the mean. 
The mean speed shown in Fig.\ \ref{fig:horVel} for each bin is marked in the 2D histogram in Fig.\ \ref{fig:SS_colorsAll_hist} for comparison.
For many measured speeds in the weaker-field bins, the advection speed is much faster (up to $\sim$700 m/s) than the bin's average speed (shown with black diamonds). We believe that the %systematic 
horizontal streaks
in the panels are due to clumps of flux elements having a wide range of field strength moving with similar speed. For example, this behavior can be observed in the region in the white box in the top right panel of Fig.\ \ref{fig:arrowsPlot}. A close-up of this region is shown in the Appendix (in Fig. \ref{fig:magnifiedImg}).

\begin{figure*}%[H] %[!htbp]
    \centering
      \includegraphics[trim={7cm 4cm 3cm 0cm},clip,scale=0.24]{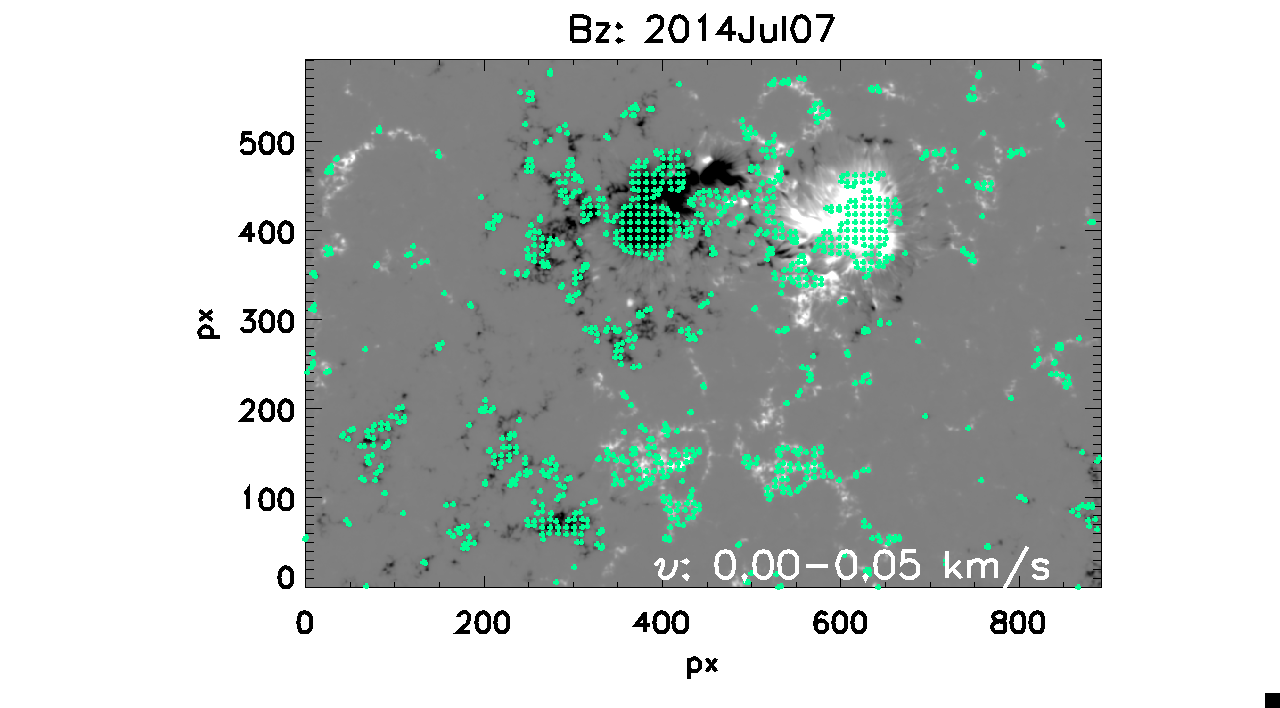}
        \includegraphics[trim={4.1cm 5.5cm 2cm 0cm},clip,scale=0.24]{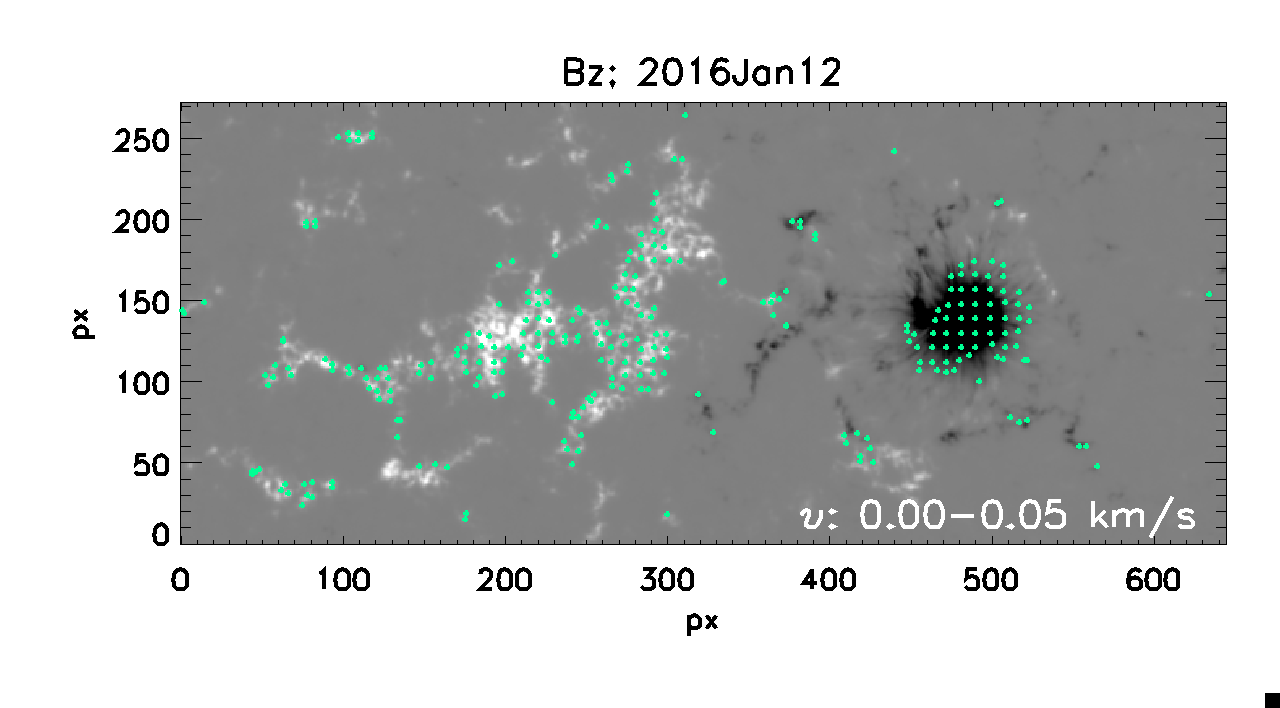}
      
      \includegraphics[trim={7cm 4cm 3cm 3cm},clip,scale=0.24]{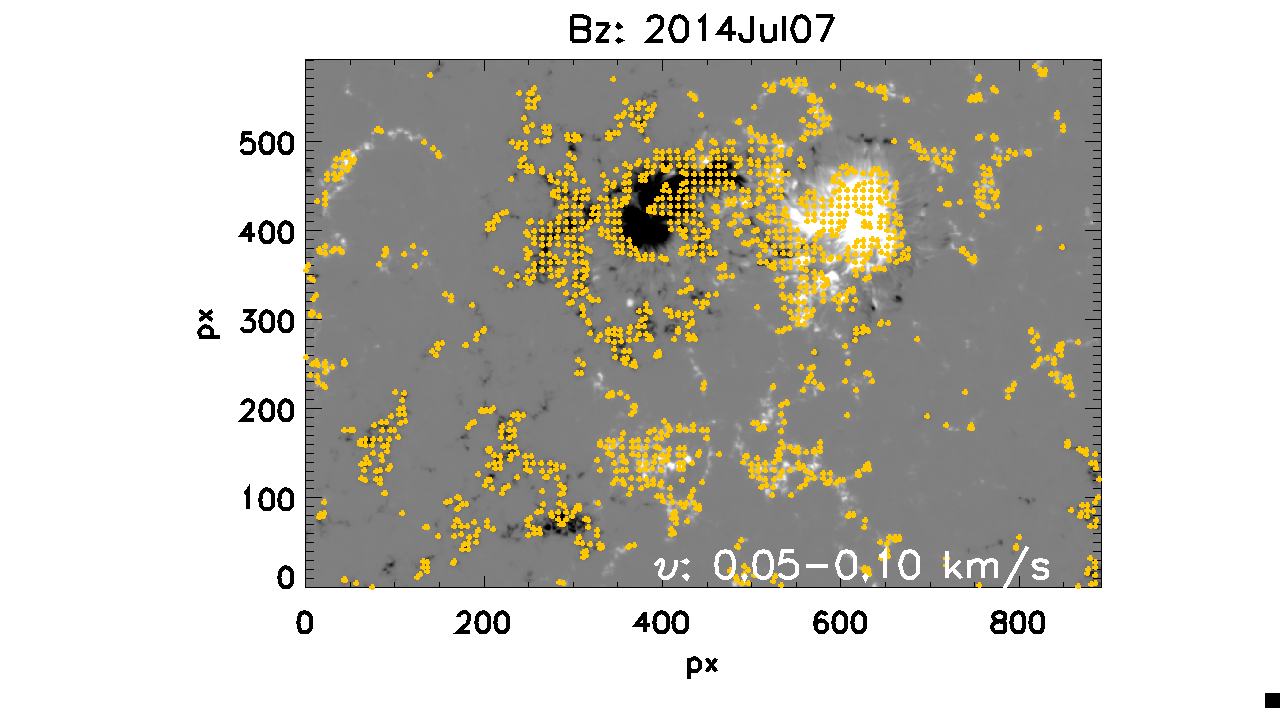}
        \includegraphics[trim={4.1cm 5.5cm 2cm 3.2cm},clip,scale=0.24]{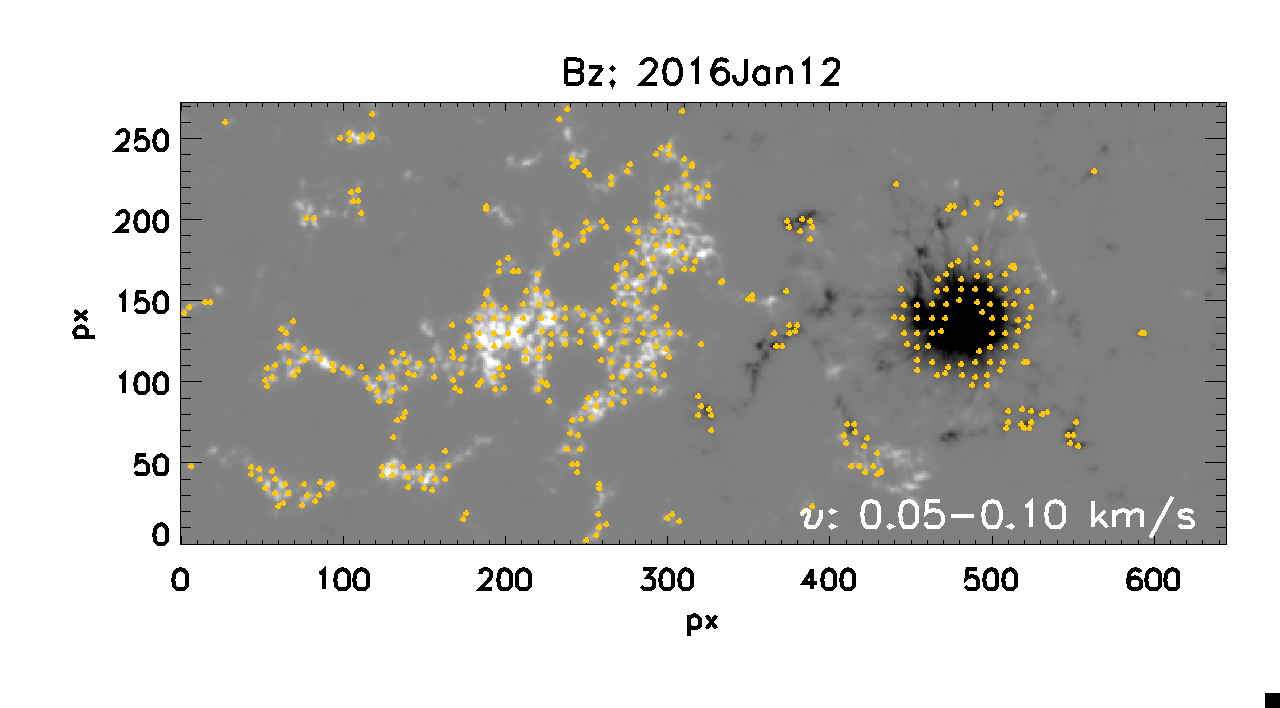}
             
      \includegraphics[trim={7cm 4cm 3cm 3cm},clip,scale=0.24]{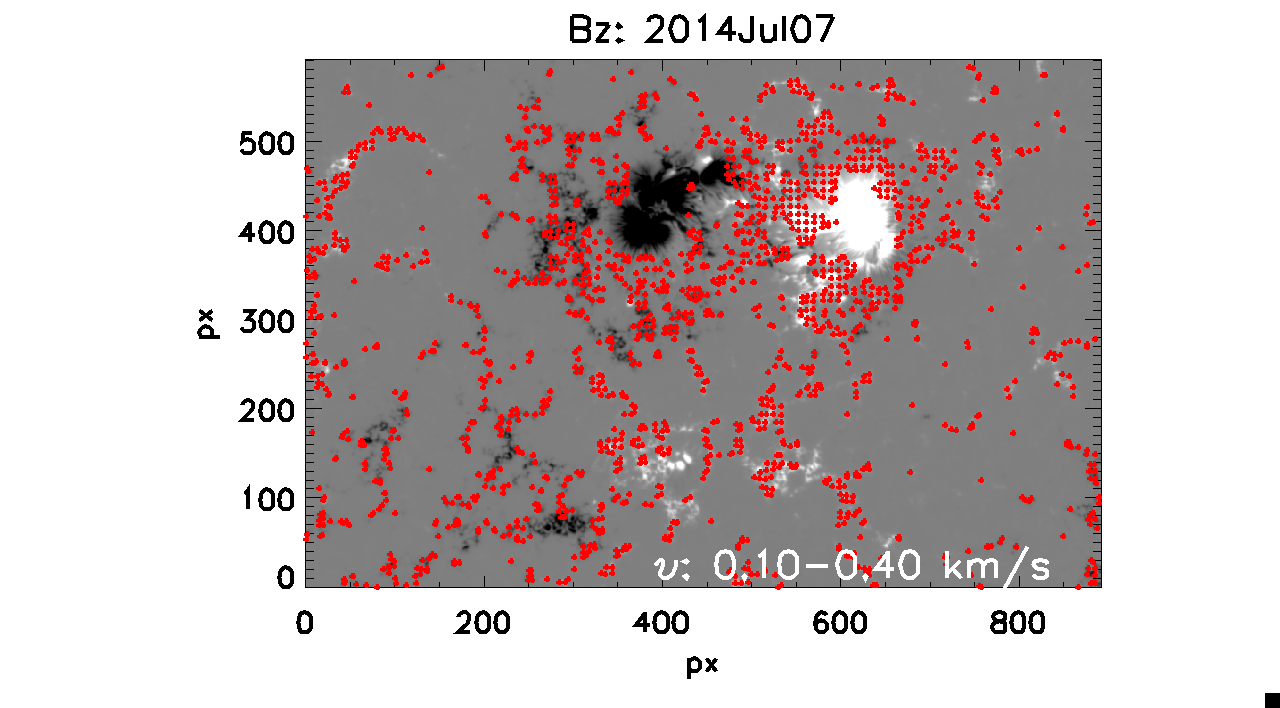}
      \includegraphics[trim={4.1cm 5.5cm 2cm 3.2cm},clip,scale=0.24]{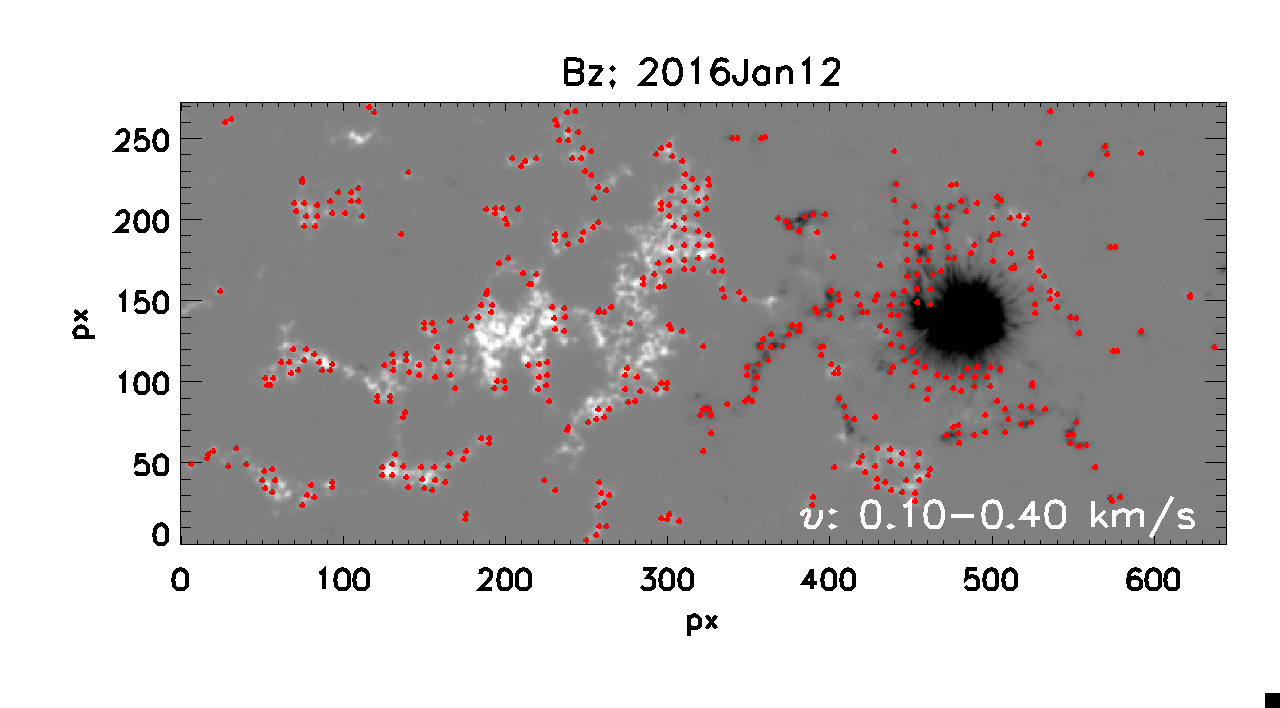}
      
       \includegraphics[trim={7cm 1cm 3cm 3cm},clip,scale=0.24]{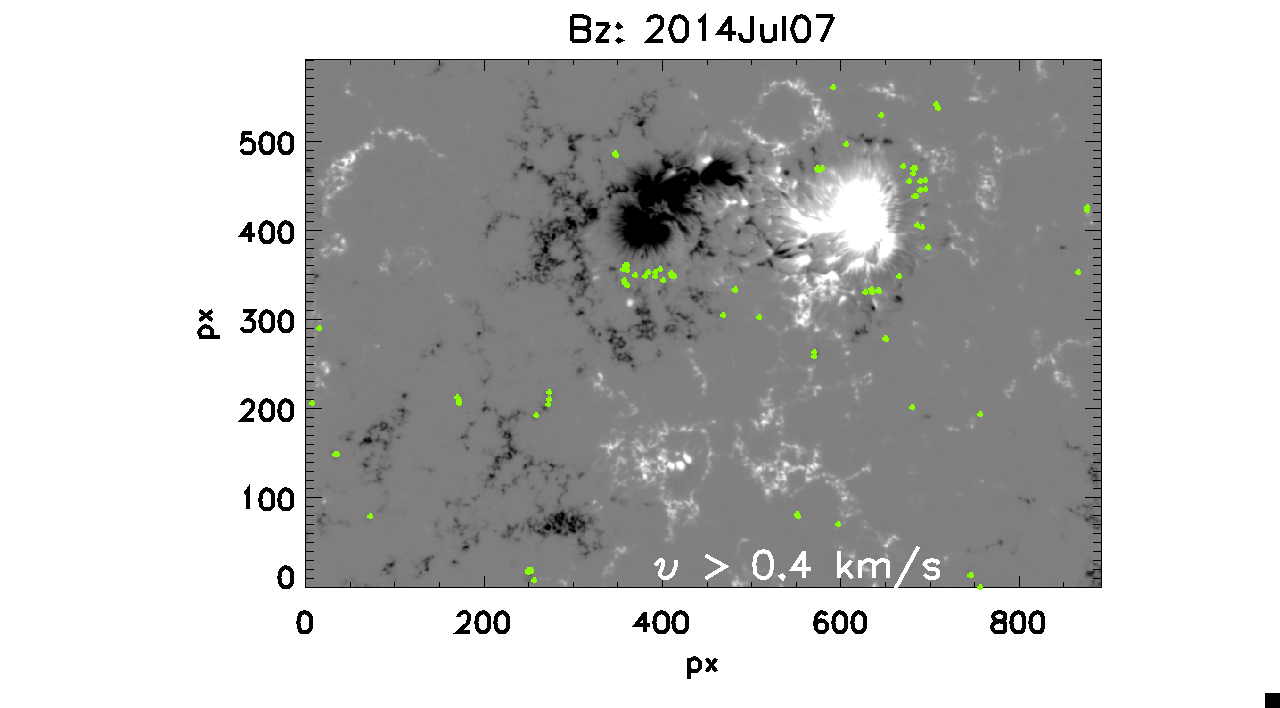}
       \includegraphics[trim={4.1cm 1cm 2cm 3.2cm},clip,scale=0.24]{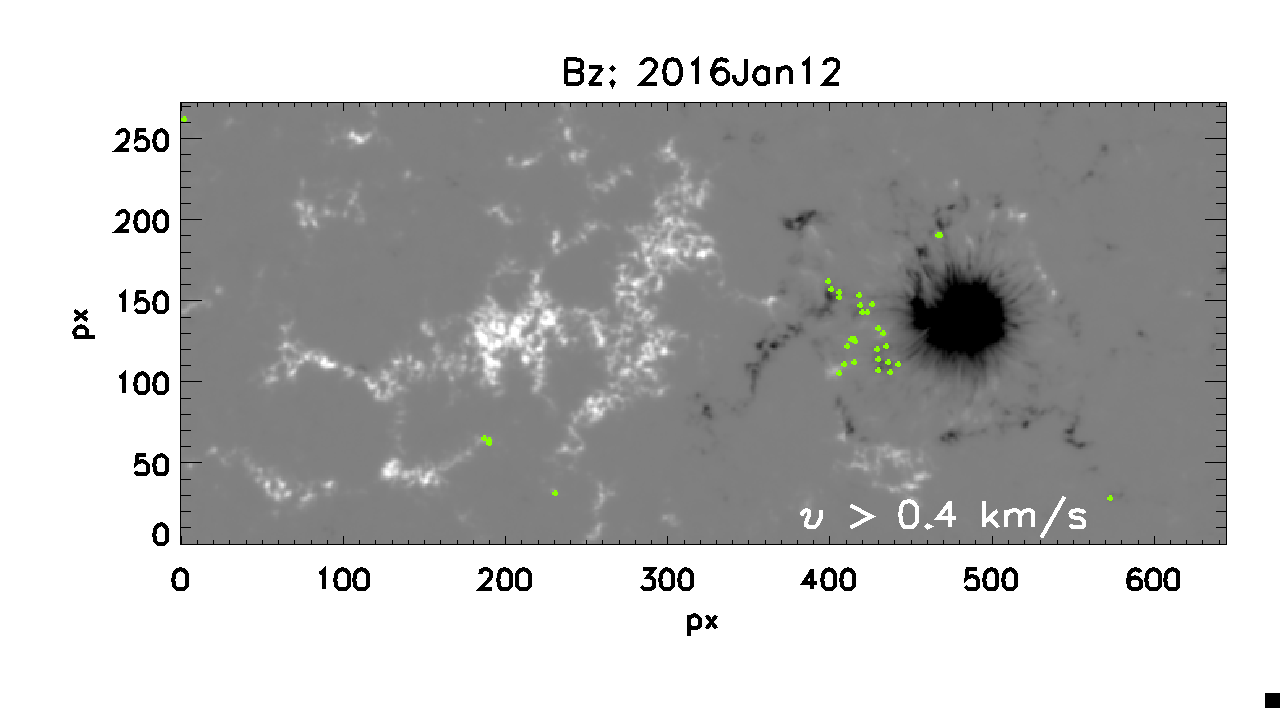}

 \caption{Average locations (centroids, colored dots) of the pixels in 9 x 9 pixel blocks for the 2-hour duration in the respective speed bins (shown in the legend in each panel) for one SS AR (left column) and one SP AR (right column). 
 The centroids are obtained as follows. All of the pixels in each speed range are marked over one common image (see text) and the centroids of those pixels in 9 x 9 pixel blocks are marked to get the panels shown here. 
 The figure essentially depicts that slow speeds ($<$ 0.05 km/s) are observed in strong field umbral regions, and in addition they may be observed outside the umbra; but, faster speeds ($>$ 0.1 km/s) are only observed outside the umbra. } \label{fig:selColLocs}
\end{figure*}

The mean locations of different speeds corresponding to features such as plage regions, penumbrae, and umbrae are of particular interest. 
These are shown in the panels of 
Fig.\ \ref{fig:selColLocs}, with the speed ranges indicated in each panel's legend. The marked mean locations are obtained by first finding all the pixels within each speed range over a two-hour interval around the central observation time and creating a binary image in which 1s (white) are the pixels in the given speed range. %Fig.\ \ref{fig:binaryImg}) 
Then, the x- and y-positions in 9 x 9 pixel blocks having at least one white pixel are averaged to get their mean location (centroid) and a dot is marked at that location. An example of a binary image is shown in the Appendix (Fig.\ \ref{fig:binaryImg}) with the grid of 9x9 pixel blocks overlaid in red. In Fig.\ \ref{fig:selColLocs}, the mean-locations thus obtained are marked on the two-hour-average magnetogram of the respective AR. We average over two hours instead of the whole 24 hours as the regions would have significantly evolved in 24 hours’ time. The centroid instead of all the white pixels in each 9x9 pixel block is marked to avoid over-crowding and for visibility of the underlying magnetic features. 

It is seen in Fig.\ \ref{fig:selColLocs} that many of the slowest speeds are in the strongest field, i.e., in sunspot umbrae (first panels in mint-green) and faster speeds are in regions of weaker field strengths (e.g., network or plage - seen in the second (yellow) and third (red) panels). Slow speeds (mint-green) can be seen outside the umbrae as well; this is typical of quiet/network region fluxes having a large standard deviation in speeds (see Fig.\ \ref{fig:horVel}). But, fast speeds are not observed in the umbra (gaps in the red panels). In the fourth panels (lime-green), regions having the highest-speeds are marked. These are in moats beyond the penumbra where there are fast moving magnetic features (MMFs; \citealt{2019ApJ...876..129L,2024A&A...686A..75Z}) that have been observed to have speeds up to 2 km/s.

\subsection{Average Advection Speed as a Function of Magnetic Field Strength}    \label{sec:FinalPlot}  
\begin{figure*}%[H] %[!htbp]
    \centering
    \includegraphics[trim={0 0.5cm 0 0},clip,scale=0.32]{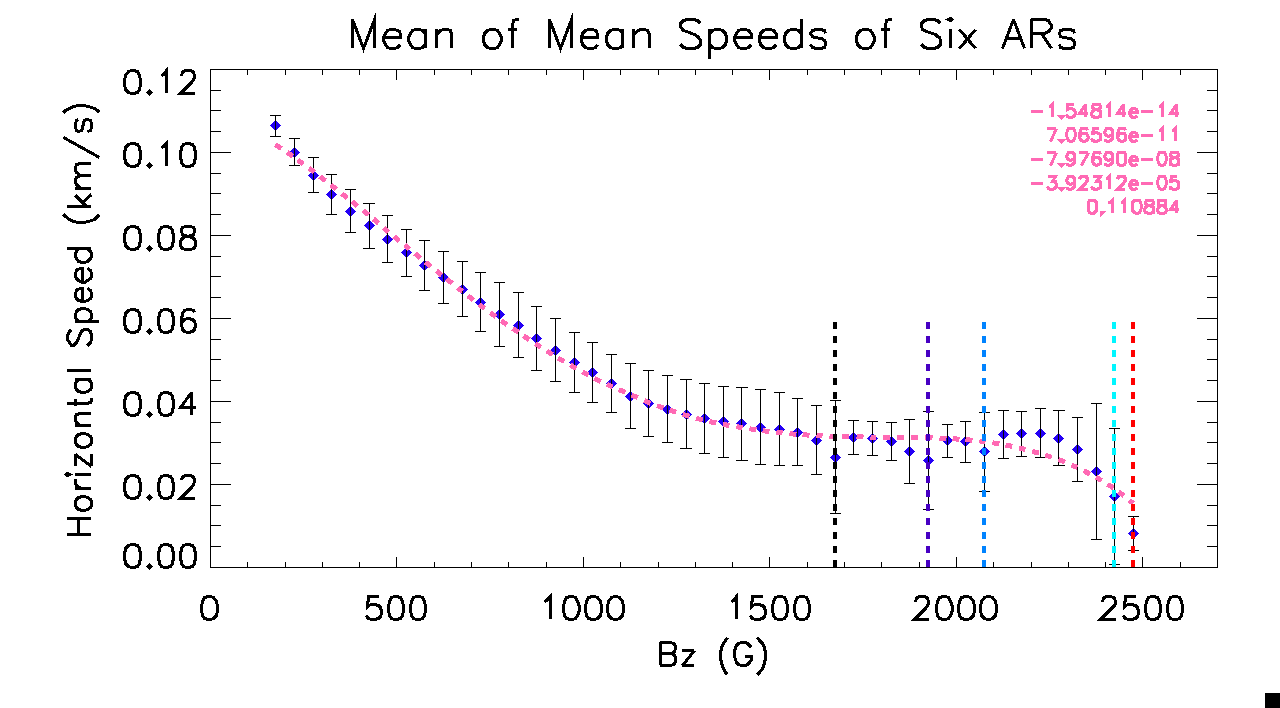}
 \includegraphics[trim={0 0.5cm 0 0},clip,scale=0.32]{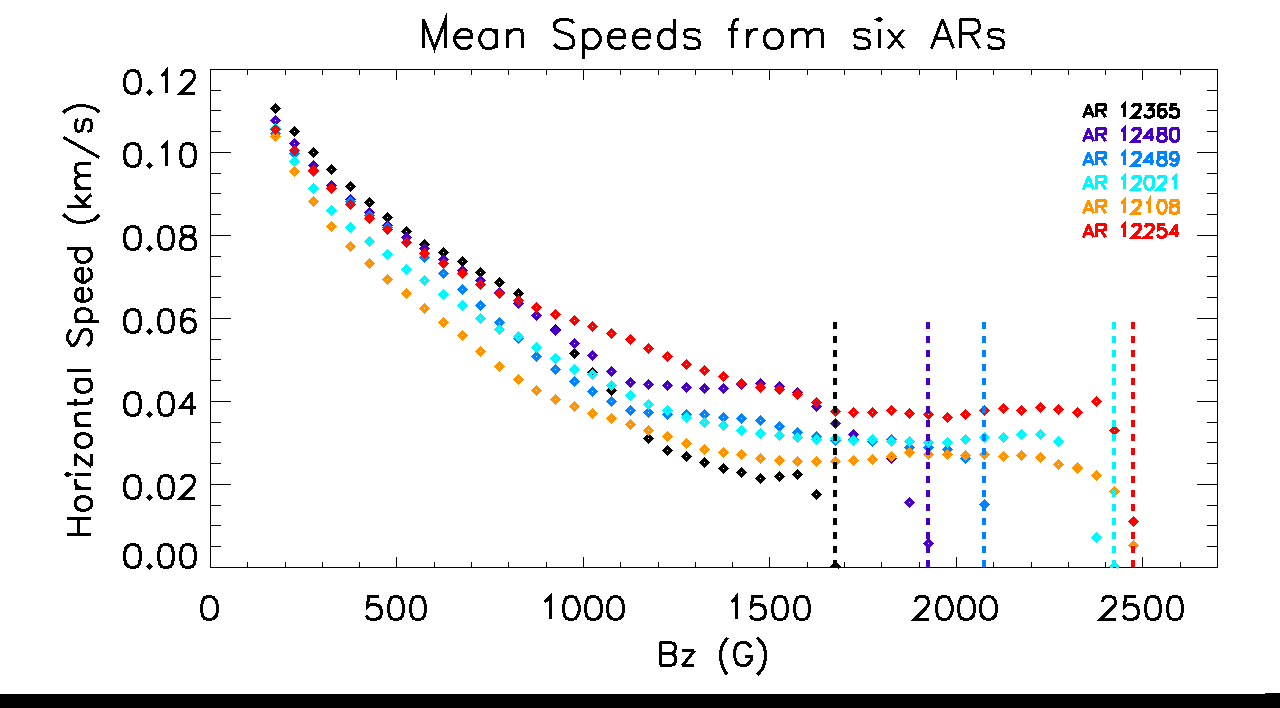}
    \caption{Top panel: Mean of the mean speeds in Bz bins for the six ARs (mean of the plots shown in Fig.\ \ref{fig:horVel}) measured in this study (blue diamonds). The error-bars (black) show the standard deviation of the values around the mean of means for each bin.  
    The pink dashed line is the fourth degree polynomial (shown in Eqn 1) fit to the means. The coefficients of the fit are given in the legend on the top right in descending order of exponents. 
    Bottom panel: Mean-speed curves for all the six ARs with the curve for each AR represented in different colored diamonds. The NOAA numbers of the ARs are given in the respective color in the legend in the bottom panel. 
     The vertical dashed line of each color marks the strongest Bz bin for the AR of that color.} 
    \label{fig:avgOfMeanVels}
\end{figure*}

To express the general trend observed in the six active regions, we take the average of the mean speeds from all the six ARs. This is shown in Fig.\ \ref{fig:avgOfMeanVels}. It can be seen that after a steady decrease in the speeds between $\sim$200-1200G (e.g., marked by trend line a.\ in the first panel of Fig \ref{fig:horVel}), it levels off despite the increasing field strength. This plateau (marked by trend line b.) continues until the curve reaches a `knee' where the mean speed rises slightly before dropping to near zero speeds at the strongest field strengths (trend line c.). 

In order to express the decline of the horizontal advection speed of the magnetic flux as a continuous function of increasing field strength, we fit a fourth degree polynomial to the mean of the mean horizontal speeds obtained bin-wise for each of the six active regions (Fig.\ \ref{fig:avgOfMeanVels}). The fit is the pink dashed curve in the top panel.
The fit is of the following form, and gives the speed in km/s, 
 \begin{equation} 
 \resizebox{\columnwidth}{!}    
  {$
    v_{h} = -1.55\times\text{10}^{-14}x^4+7.06\times\text{10}^{-11}x^3-7.97\times \text{10}^{-8}x^2-3.92\times\text{10}^{-5}x+0.11
    $}
\end{equation}\label{eqn1}

\noindent Here $x$ is the value of the vertical magnetic field (Bz). 
In Fig.\ \ref{fig:avgOfMeanVels}, the error-bars for the blue diamonds are the standard deviations around the mean horizontal speeds from the six ARs. The coefficients corresponding to the 4th-degree polynomial are also given in the legend in the plot in the top panel. Higher order fits fit more perfectly, but this curve provides a reasonable estimate of the flux advection speed as a function of Bz in this range of Bz on the solar surface. 

\subsection{Speeds Across Penumbra and Umbra}\label{velCut_umbraPenumbra}
\begin{figure*}%[H] %[!htbp]
\centering
\includegraphics[trim={1cm 3cm 3cm 0cm},clip,scale=0.22]{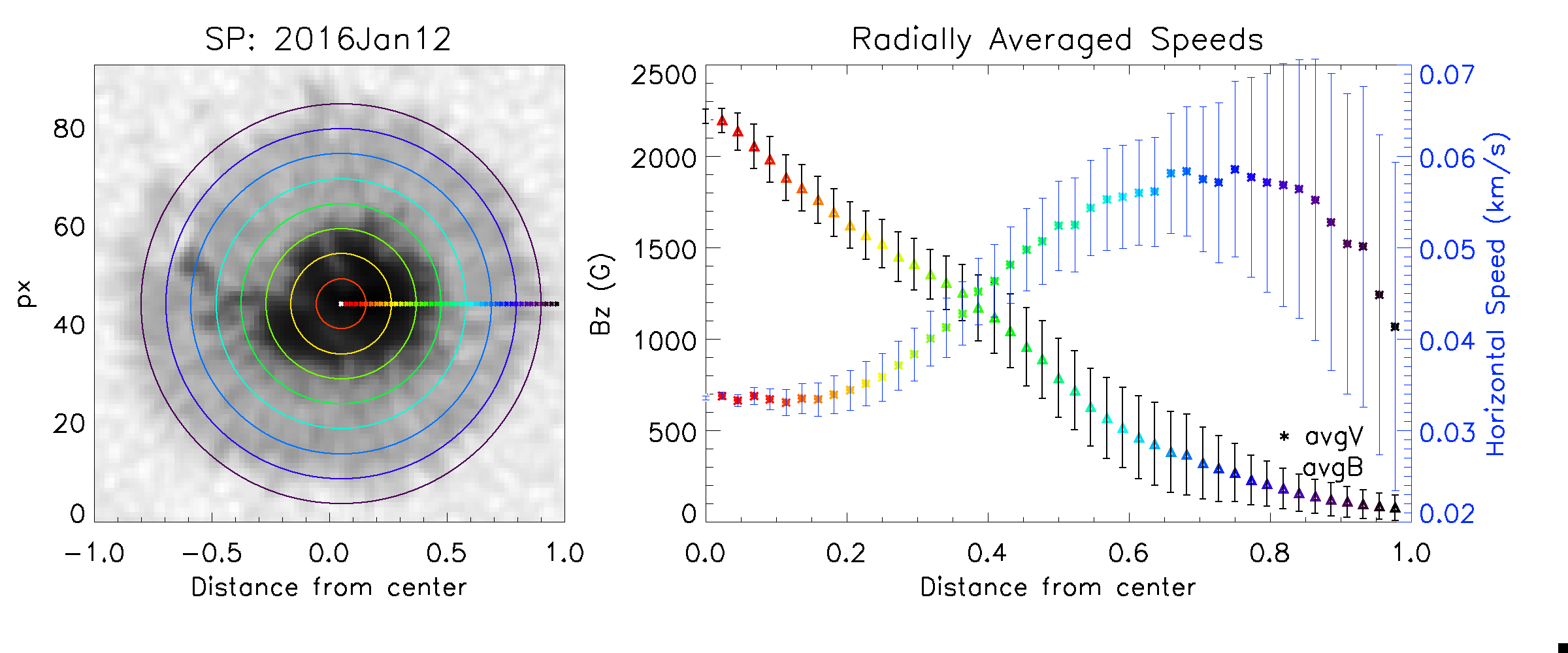}
\caption{ 
Left Panel: Continuum image of the sunspot in AR 12480 (2016Jan12). 
Right panel: Horizontal speed averaged (asterisks) on 1-pixel thick concentric circles over two hours from the center to the outer edge of the sunspot region shown in the left panel. Each average field strength (triangles) from center to the outer edge of the sunspot has the same color as the average speed for that circle. The standard deviations around the averages are shown in black thick and blue thin error bars for magnetic field strength and speeds, respectively. The standard deviations for the speeds (blue error bars) are divided by four to better display them. The standard deviations for the magnetic field strength (black error bars) have their true size. }
\label{fig:penumum_cut} 
\end{figure*}

Figure \ref{fig:horVel} shows that the mean advection speed decreases smoothly with increasing vertical
field strength with a plateau (marked by trend line b.\ in the top panel of Fig.\ \ref{fig:horVel}) at strong fields. 
In these mean-speed trends, there is no abrupt change in the speeds at the penumbra-umbra boundary, which seems contrary to the expectation that the speeds are higher in the inner penumbra and lower right inside the umbra. To explore this and provide evidence that due to averaging the speeds in each bin we do not see a sharp decline in horizontal speed at the Bz typical of the umbra-penumbra boundary, we average the speeds radially on concentric circles (see left panel in Fig.\ \ref{fig:penumum_cut}) of increasing radius by one pixel to obtain the average speeds with average field strength from the center of the umbra to the outer edges of the penumbra. The average speeds (in asterisks) and corresponding average field strengths (in triangles) with increasing distance from the center of the umbra are shown in the right panel of Fig.\ \ref{fig:penumum_cut}. The colors correspond to the respective concentric circles in the sunspot are separated by five pixels in the left panel to avoid overcrowding. 
We chose this sunspot for this demonstration because it is almost round and has radially symmetrical penumbra. 
As expected, the average flux advection speeds (asterisks) are slowest in the umbra (strongest fields, represented by triangles of the same color) than in the penumbra and lower back down outside the penumbra. The decrease in horizontal advection speeds at and beyond the sunspot’s edge arises from the reduced density of flux outflow patches \citep{2013A&A...557A..24V,2013A&A...557A..25T,2015A&A...583A.119T}. Moreover, the large variability in flux advection speed at these radii reflects the differing horizontal speeds of outflowing strong-field features such as moving magnetic features \citep{2007A&A...471.1035Z,2019PASJ...71R...1H}. 

The horizontal advection speeds of flux in and around the sunspot in Fig.\ \ref{fig:penumum_cut} (including the range of the standard deviation) range between 20 and 300 m/s consistent with previous studies \citep{2013SoPh..287..279L,2022A&A...662A..13S}. We conclude that the plateau and lack of abrupt change from umbra-penumbra boundaries in Fig.\ \ref{fig:horVel} result from the averaging of this broad range of speed from across the penumbra-umbra edge, and from the flux of the same Bz in plage and network regions. 

%%%%%%%%%%%%%%%%%%%%%%%%%%%%%%%%%%%%%%%%%%%%%%%%%%%%%%%%%%%

\section{Summary and Discussion} \label{sec:disc}

We have applied FLCT on six active regions - three of them having leading and trailing sunspots (SS) and three having leading sunspot and trailing plage (SP). We have mainly used SHARP CEA Bz magnetograms for characterizing the effect of magnetic field strength on horizontal advection speed. 
Past studies have analyzed the plasma motions in specific regions on the solar surface such as quiet regions, plage, or sunspot regions, with different types of data. This work includes all these regions by choosing non-flaring active regions (ARs) surrounded by both plage and quiet regions, thus covering a wide range of field strength simultaneously using HMI SHARP Bz magnetograms. To our knowledge, this is the first study that uses magnetograms with a wide range of field strength to quantify magnetic flux advection speed as a function of vertical field strength. 

It is evident from Fig.\ \ref{fig:avgOfMeanVels} that, in general, as the vertical field strength gradually increases, the horizontal speed decreases, indicating that stronger vertical fields increasingly suppress the horizontal surface motions. The movies of each AR with the horizontal velocity arrows overlaid (available in the online version) also show this trend clearly, where longer arrows are seen in relatively quieter regions, intermediate length arrows in plage and penumbral regions, and shortest arrows in the sunspot umbrae. 
One frame from the animation for each AR with the velocity arrows overlaid on the magnetograms are shown in Fig.\ \ref{fig:arrowsPlot}. Note that a threshold of 150 G is used with FLCT, hence the arrows are obtained only for regions where field strengths exceed this value, excluding weaker flux in quiet regions. That means that the quiet-Sun velocities we obtain are not for the weaker flux, and the weaker flux less than 150 G may have much higher horizontal speeds. 

\cite{1992ApJ...393..782T} were the first to demonstrate the relation between field strength and horizontal advection speed by using LCT on continuum intensity images and obtaining the magnetic field in the corresponding locations from co-aligned, co-temporal magnetograms, focusing on a plage region. Their result is shown in Fig.\ 27 of their paper. It is important to note that whereas they used LCT on continuum intensity images, we use FLCT on magnetograms. Although their magnetic field strengths only reach up to about $\sim$800G (plage regions), the overall trend of the horizontal speed vs.\ magnetic field strength aligns with our result. The cadence of their images and magnetograms is $\sim$60sec, and their measured speeds ranged from 370 m/s in weakest-field regions to 125 m/s in plage. Because of the faster cadence, their measurements tracked features that are faster and shorter-lived compared to ours. We obtain comparable speeds with BLOS data at 45 sec cadence (see Fig.\ \ref{fig:horVel_BzBlos} in Appendix). 

\cite{2012A&A...537A..85S} applied LCT on a pore region of AR 11005 using IBIS (Interferometric Bidimensional Spectrometer, \citealt{2006SoPh..236..415C}) white light images from the Dunn Solar Telescope to measure the horizontal and LOS velocities in and around the pore and correlated them with the field strength there. Their LOS field strength ranges from zero in the quiet Sun around the pore to about 1500 G in the core. Similar to the results of our work, they find a one-to-one correlation with a decrease in the horizontal and LOS velocities as BLOS increases. 
The horizontal speeds they obtained from BLOS magnetograms is similar to those we obtain using the 45-second BLOS magnetograms. 

The study by \cite{2013SoPh..287..279L} is, to our knowledge, the only prior work that used an optical flow method (using DAVE4VM, \citealt{2008ApJ...683.1134S}) on 12-minute vector-magnetograms (Bx, By, Bz) from SHARP to compare the photospheric surface velocities with subsurface flows in a sunspot region, with a Gaussian kernel size of 19 pixels similar to ours (15 pixels). The sunspot they use (AR 11084) is round and similar to our 2016Jan12 spot (AR 12480). They found that the photospheric flows are inward inside the umbra and outward in the penumbra, as quantified by \cite{2022A&A...662A..13S}. Although they did not specify the speeds in various regions in and around the spot, the flow speeds they observe do not appear to exceed those that we measure in strong field regions using the 12-minute Bz dataset. Unlike their study, we use only Bz magnetograms because of the broader range of field strengths we examine, as the noise from the Bx and By vector data is higher. This is higher especially at lower field strengths, where the noise from the transverse field components (Bx, By) dominate the signal, which makes it difficult to obtain reliable speed values. Additionally, for ARs located near the disk center, except in penumbra, the Bz component is considerably larger than the transverse components of the field. All the ARs used in our study are located well within 30$^{\circ}$ of the disk center, making SHARP Bz data the most suitable for our analysis. 

The overall lower speeds in our horizontal speed versus Bz curves (Fig.\  \ref{fig:horVel} and Fig.\ \ref{fig:avgOfMeanVels}), compared to \citet{1992ApJ...393..782T} suggest that slower features are tracked at lower cadences (longer interval between images), with the Bz data cadence being 12-minute. This interpretation is further supported by the gradual reduction in the velocities seen in BLOS data between 45-second, 3-minute, 6-minute and 12-minute cadences, as shown in Fig.\ \ref{fig:horVel_BzBlos} (also see Figs.\ 13 and 14 in \citealt{2012ApJ...747..130W}).
At higher field strength bins in the 45-second curves, the average speeds are higher. Remnants of these appear to be present as bumps near the highest field strengths in 3, 6 and 12-minutes BLOS curves as well. We suspect these to be due to p-mode oscillations and might be important for heating of coronal loops stemming from penumbra, although loops connecting opposite-polarity sunspot umbrae have not been observed in cool, warm or hot channels \citep{1979ApJ...229..375P,1981SoPh...69...99W,2017ApJ...843L..20T}.

As mentioned in the previous section, the horizontal speed versus Bz trend can be divided into a steady-decrease, plateau and knee parts (shown with trend lines a.,b.,c.\ in top left panel of Fig.\ \ref{fig:horVel}). The plateau range in the plots likely comes from the outer umbra and the inner penumbra, where the highly variable flux advection speeds from penumbral filaments 
are averaged in. The plateaus may also include advection speeds from moats or MMFs and advection speeds due to umbral and penumbral waves.

{In Fig.\ \ref{fig:penumum_cut}, the reduction in the speed outside the penumbra possibly results from the reduced number of high-speed features such as the MMFs.} {In sunspot umbrae, the horizontal velocities of peripheral umbral dots, which are known to typically move inward towards the umbral core, have measured speeds up to 2.0 km/s \citep{2009ApJ...702.1048W,2012ApJ...745..163K}. The speeds that we measure in the umbra are much smaller ($\sim$0.02 km/s), because we measure the speeds of magnetic flux elements, whereas umbral dots might be field-free regions. }

The mean speeds in Fig.\ \ref{fig:horVel} and Fig.\ \ref{fig:avgOfMeanVels} increase slightly after the plateau and form a slight bump before dropping off to near zero (line trend c.\ in Fig.\ \ref{fig:horVel}) at the highest field strengths. We speculate that this bump is possibly due to the penumbral/ umbral p-mode oscillations. 
Strong field regions such as the umbra are known to suppress convection and the p-mode oscillations \citep{1977ApJ...213..900B}. Magneto-acoustic waves propagate along inclined magnetic field lines \citep{2004Natur.430..536D,2006ApJ...647L..77M}, which are prevalent in the outer umbra and penumbra \citep{1992ApJ...394L..65B}. The horizontal speeds in the bump beyond the umbral line %(red lines in Fig.\ \ref{fig:horVel}) 
are possibly due to Bz changes due to these oscillations. Advection of strong magnetic flux in moats might also contribute to these speeds \citep{2007A&A...472..277S}. 
Going further into the strongest fields in the umbra, the horizontal speeds tend to zero indicating not only strongest suppression of advection, but also the {presence of mostly} vertical fields in which p-mode oscillations are not detected.

It should be noted that we have focused here on the relationships between flow speeds and vertical and line-of-sight magnetic field strengths, but we have not presented the relationship between flow speeds and the magnitude of the full vector {magnetic field}, or the magnitude of the horizontal field component. Characterizing the relationships between flow speeds, whether inferred by tracking Bz or other methods, and either total or horizontal field strengths could further illuminate magnetic influences on the Sun's near-surface convection. Also, we expect that convective flows exist over a range of spatial scales, and more power in flows at scales accessible to tracking might be correlated with stronger unresolved flows that are believed to produce line broadening. If so, then we would expect a correlation between tracking speeds and line widths inferred by the HMI inversion pipeline used to produce the magnetograms, inspecting which is out of the scope of the present work.

In addition to obtaining a quantitative relation between magnetic field strength and advection speed, this work is important in the context of coronal heating. Wave heating and magneto-convection are the primary mechanisms thought to heat the corona. Coronal heating from magneto-convection can occur due to nano-flares at tangential discontinuities in the field lines due to twisting and braiding by convection-driven wrapping of the field lines around one another \citep{1988ApJ...330..474P,2021NatAs...5..237B}.  
It is possible that the convective freedom at the loop feet can also produce waves that dissipate their energy along the loops and cause heating. However, at this time it is not possible to differentiate between the two types and to what degree they are produced or suppressed in different magnetic field strength regions and it needs further exploration.
As evident from X-ray and extreme-ultraviolet observations, not all loops are similarly bright or hot. \citet{2017ApJ...843L..20T} concluded that loops connecting a pair of sunspot umbrae are invisible due to the minimal convection within the umbrae and that both freedom of convection and magnetic field strength are important for heating and brightening coronal loops. 
Coronal heating scaling laws so far have used the magnetic field strength and the loop-length as the dependent parameters for the scaling laws of the temperature/ brightness of loops. For example, \citet{2002ApJ...571L.181B} and \citet{2019ApJ...877..129U} show such a relationship with the temperature/ brightness being directly proportional to the field strength. These laws are based on observations or modeling omitting the sunspot umbra-to-umbra loops, and fail to fit these loops. The motivation for the current work primarily stems from the above mentioned evidence from \citealt{2017ApJ...843L..20T}. The horizontal speed as the function of field strength obtained here will be used to build a more holistic scaling law that might be able to include umbra-to-umbra loops. That investigation will be presented in a subsequent paper.

%\begin{acknowledgements}
{We sincerely thank the anonymous referee for their insightful comments which have significantly improved the manuscript. SKT, VA, and RLM gratefully acknowledge the support by NASA HGI award (80NSSC21K0520). SKT, VA, NKP and RLM sincerely acknowledge support from HSR grant (80NSSC23K0093) and/or NSF AAG award (no.\ 2307505). 
NKP gratefully acknowledges support from NASA’s SDO/AIA, and HSR grant (80NSSC24K0258). BTW gratefully acknowledges support from NASA HSR-80NSSC23K0092. 
We have used data from SDO/AIA and HMI instruments. AIA is an instrument onboard the Solar Dynamics Observatory, a mission for NASA’s Living With a Star program. We acknowledge NASA ADSABS, Solar Software and Python libraries. }
%\end{acknowledgements}

%%%%%%%%%%%%%%%%%%%%%%%%%%%%%%%%%%%__________APPENDIX_________%%%%%%%%%%%%%%%%%%%%%%%%%%%%%%%%%%%%%
\appendix
%\section{Appendix}
\subsection{Correlation Analysis for Data Choice}

\begin{figure}[H] %[!htbp]
\centering
    \includegraphics[trim={2cm 0.5cm 0 0},clip,scale=0.2]{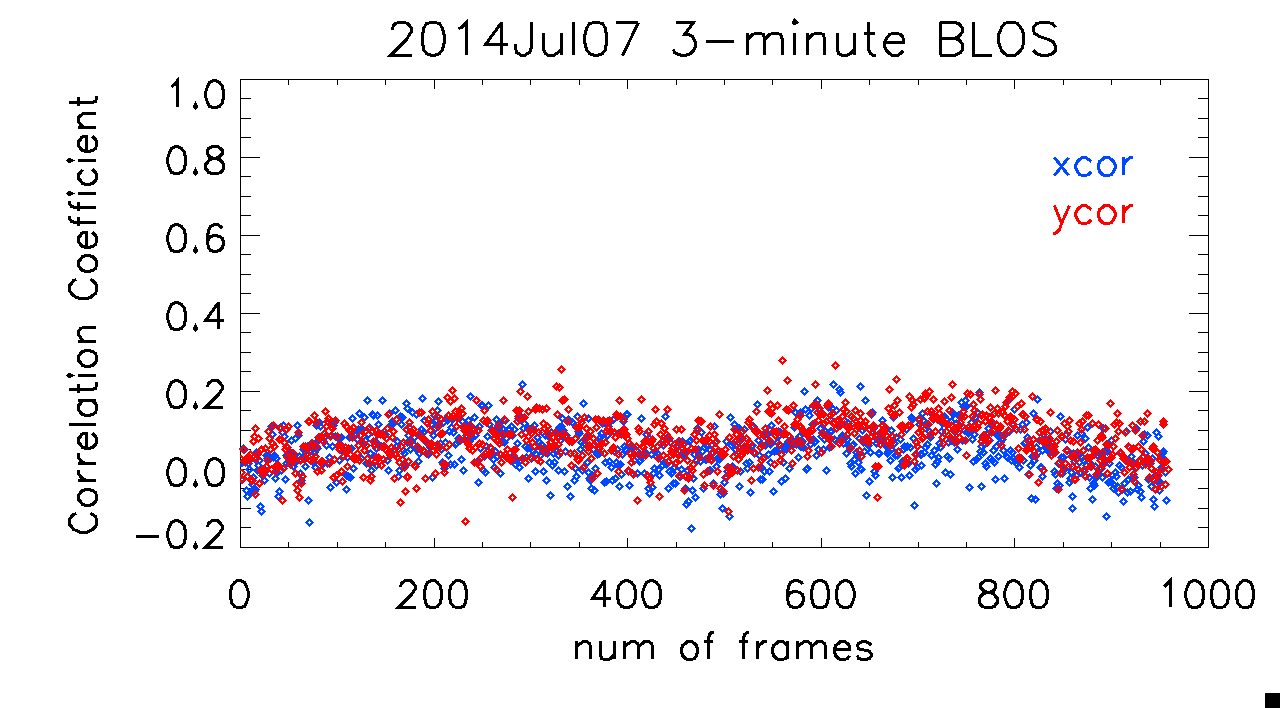}
    \includegraphics[trim={4cm 0.5cm 0 0},clip,scale=0.2]{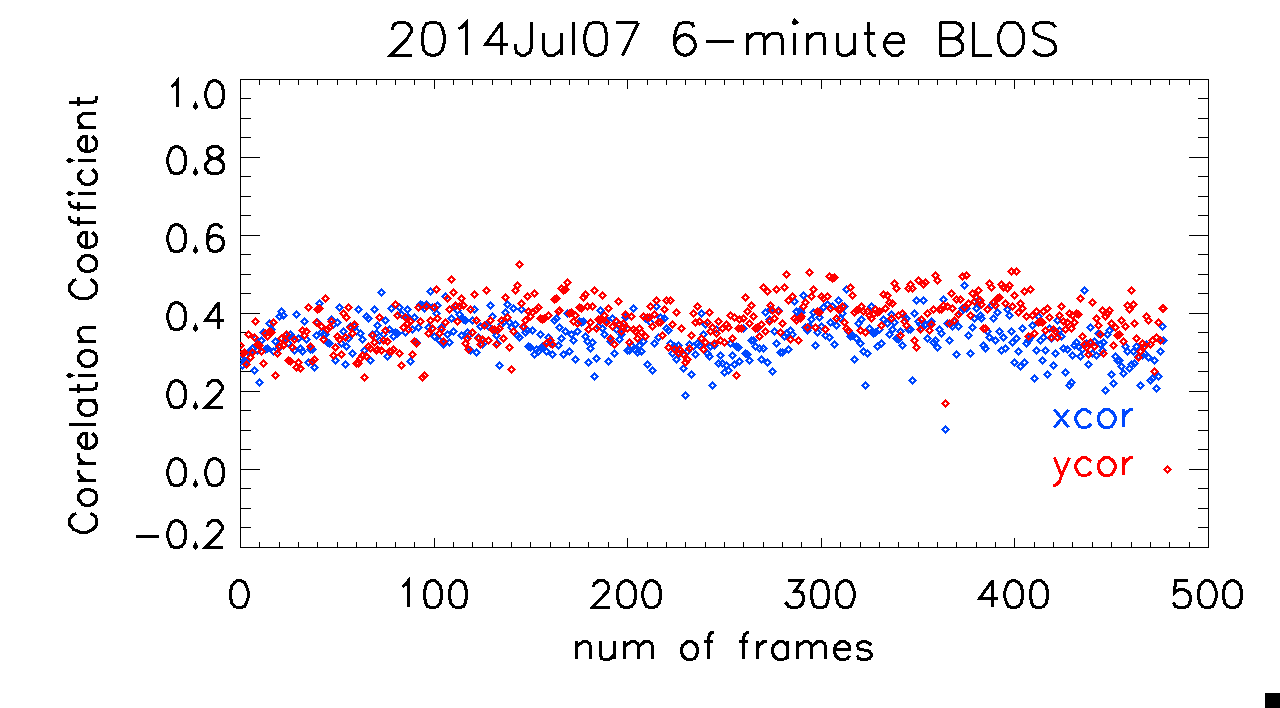}
    \includegraphics[trim={2cm 0.5cm 0 0},clip,scale=0.2]{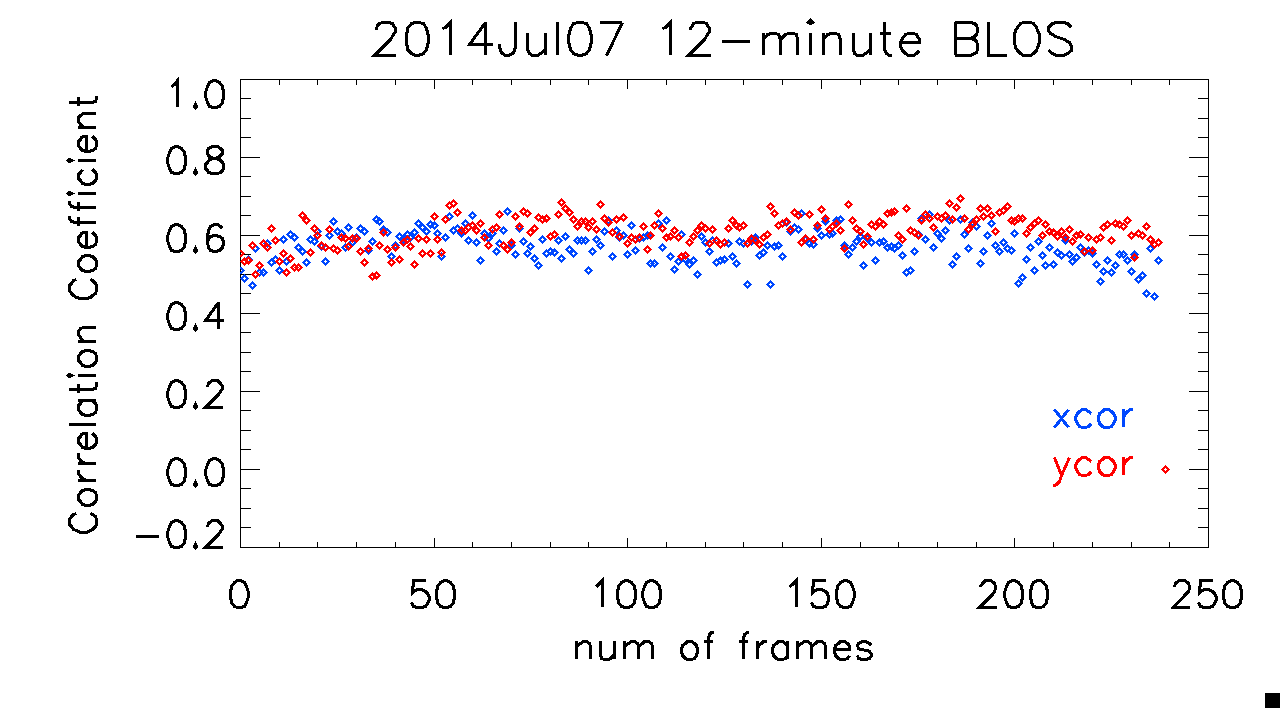}
    \includegraphics[trim={4cm 0.5cm 0 0},clip,scale=0.2]{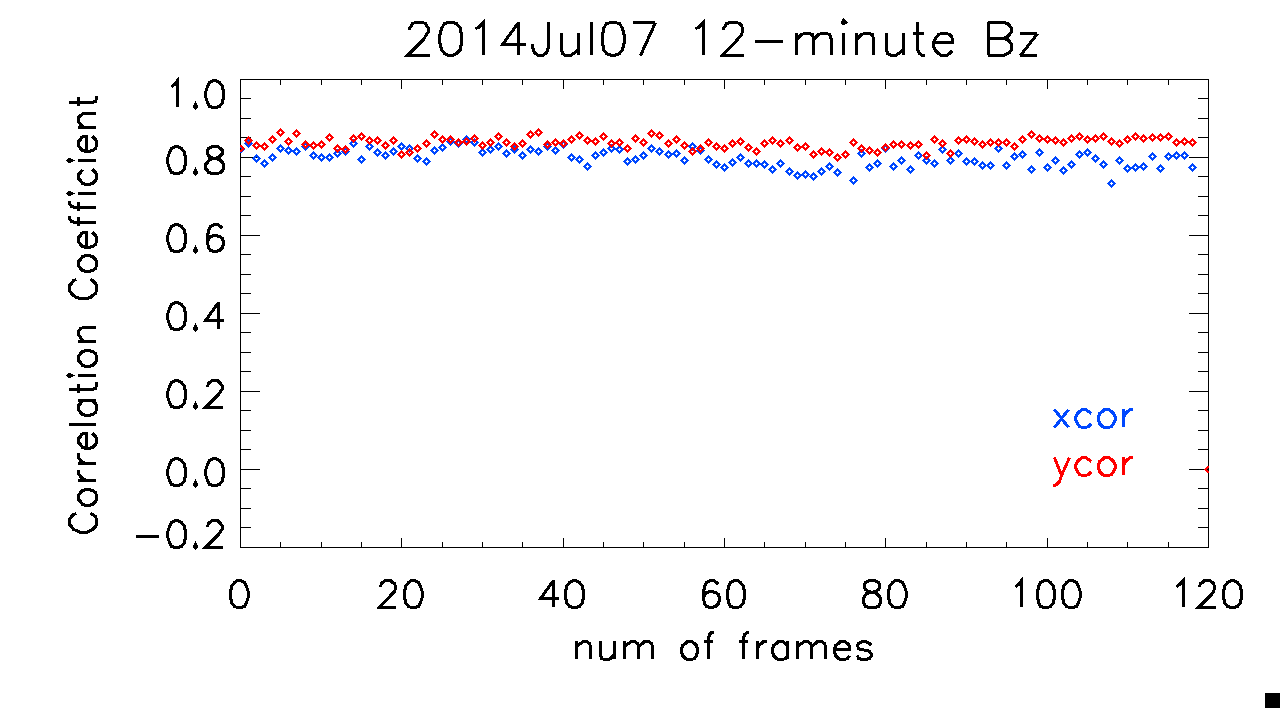}
\caption{Scatter plots of the auto-correlation coefficients of the horizontal speeds obtained using the 3-minute and 6-minute BLOS magnetograms in the top row, and the 12-minute BLOS and Bz magnetograms in the bottom row. Red diamonds show the correlation coefficients of $v_{y}$ and blue ones show those of $v_{x}$.} \label{fig:corrAnalysis}
\end{figure}

To judge which type of magnetograms to use (Bz or BLOS) and at what cadence for achieving our primary goal of obtaining the functional dependence of the horizontal advection speed on the field strength, we use the auto-correlation of advection speeds from consecutive magnetograms \citep{2012ApJ...747..130W}. 
We show auto-correlation scatterplots for 3-, 6- and 12-minute BLOS and Bz images as a gauge of the reliability of the measured speeds in Fig.\ \ref{fig:corrAnalysis}. Not shown are the plots for 36 and 60-minute cadences that showed lower correlation coefficients than those with the 12-minute Bz data.
In addition to visually examining the persistence of the velocity vectors overlaid on the magnetograms in the movies, we rely on these scatterplots. The auto-correlation coefficients of the advection speeds using HMI SHARP-CEA Bz magnetograms are larger and less scattered compared to those obtained using LOS magnetograms. Hence, we choose to use Bz magnetograms at 12-minute cadence instead of BLOS magnetograms for our analysis.
The greater stability of flows inferred from the Bz magnetograms might be due to the extensive averaging used to create this dataseries; filtergrams are averaged over a 1215 sec boxcar with a cosine weighting function with an FWHM of 720 seconds \citep{2013SoPh..287..279L,2012SoPh..275..285C}.

\subsection{Horizontal Advection Speeds using BLOS Magnetograms}

\begin{figure}[H] %[!htbp]
\centering
    \includegraphics[trim={3cm 0.5cm 1.0cm 0},clip,scale=0.20]{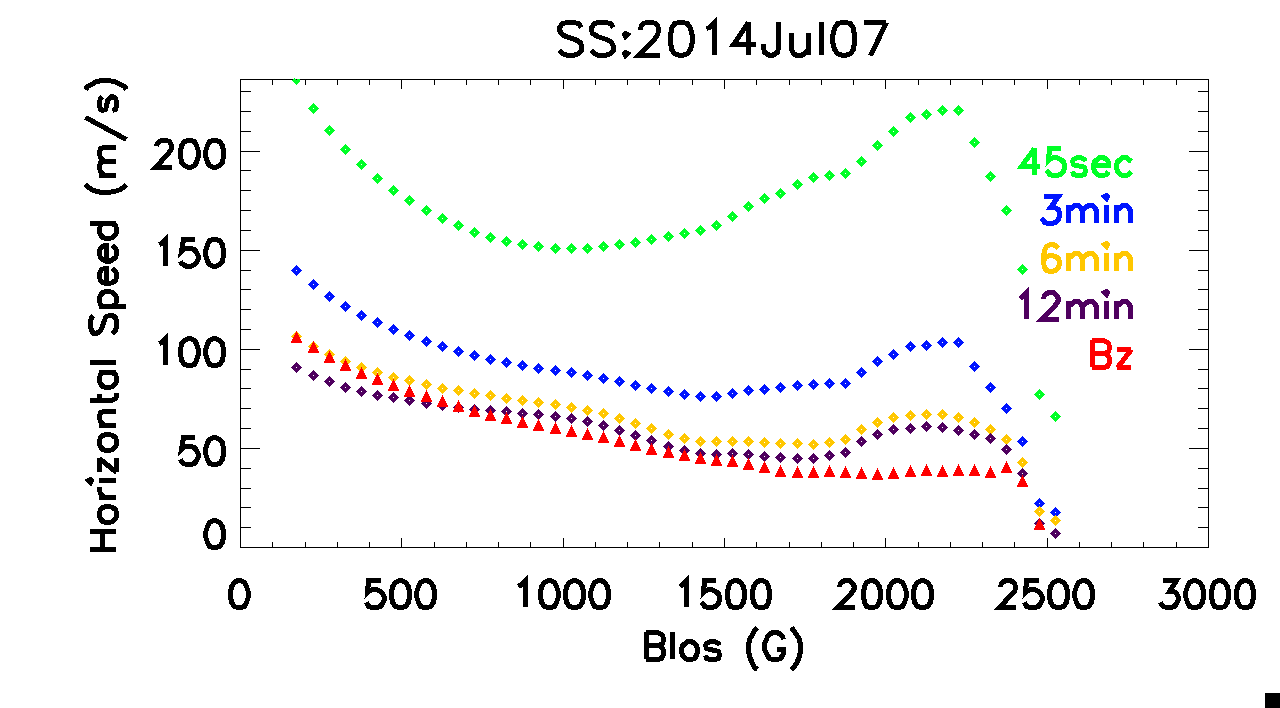}
    \includegraphics[trim={4.5cm 0.5cm 1.0cm 0},clip,scale=0.20]{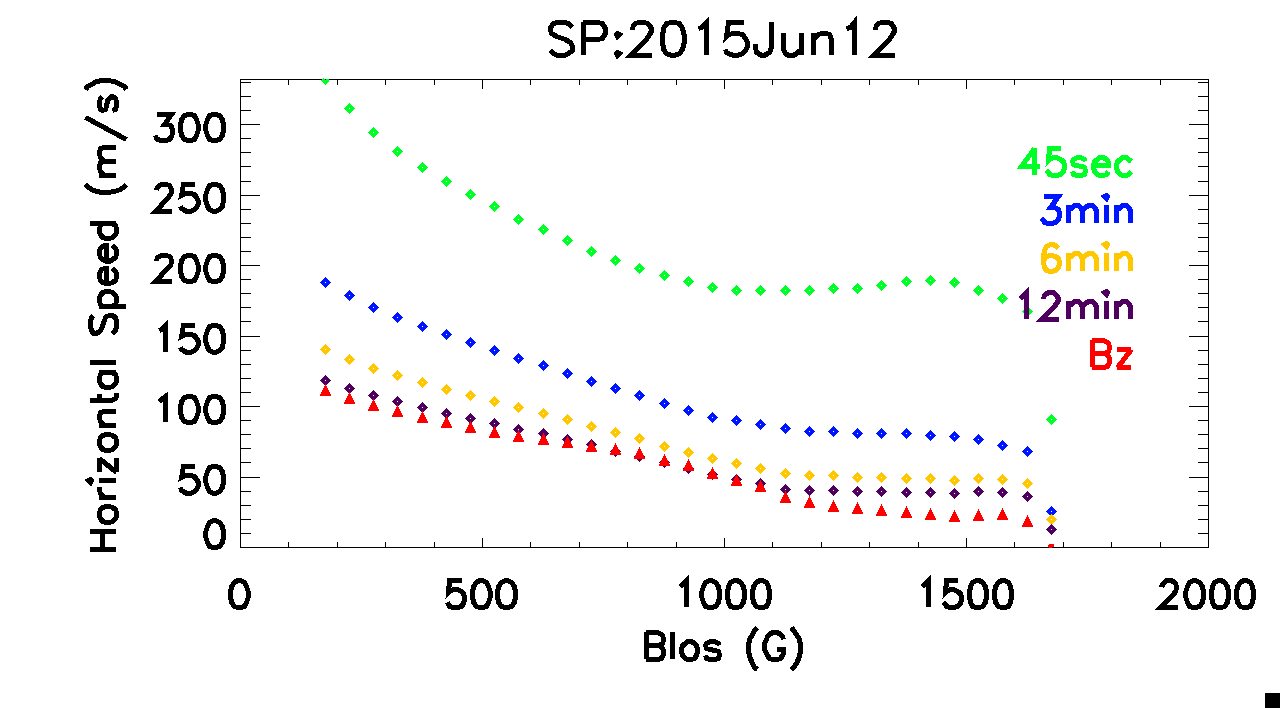}
    
    \includegraphics[trim={3cm 0.5cm 1.0cm 0},clip,scale=0.20]{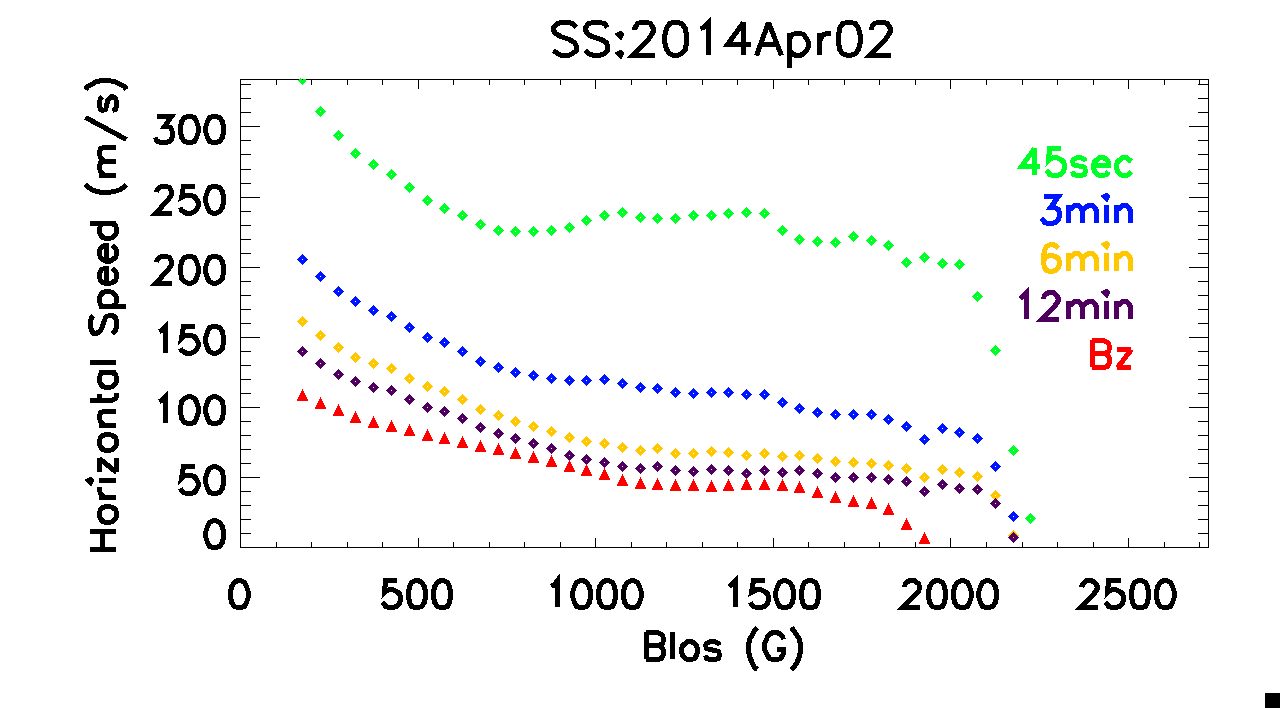}
    \includegraphics[trim={4.5cm 0.5cm 1.0cm 0},clip,scale=0.20]{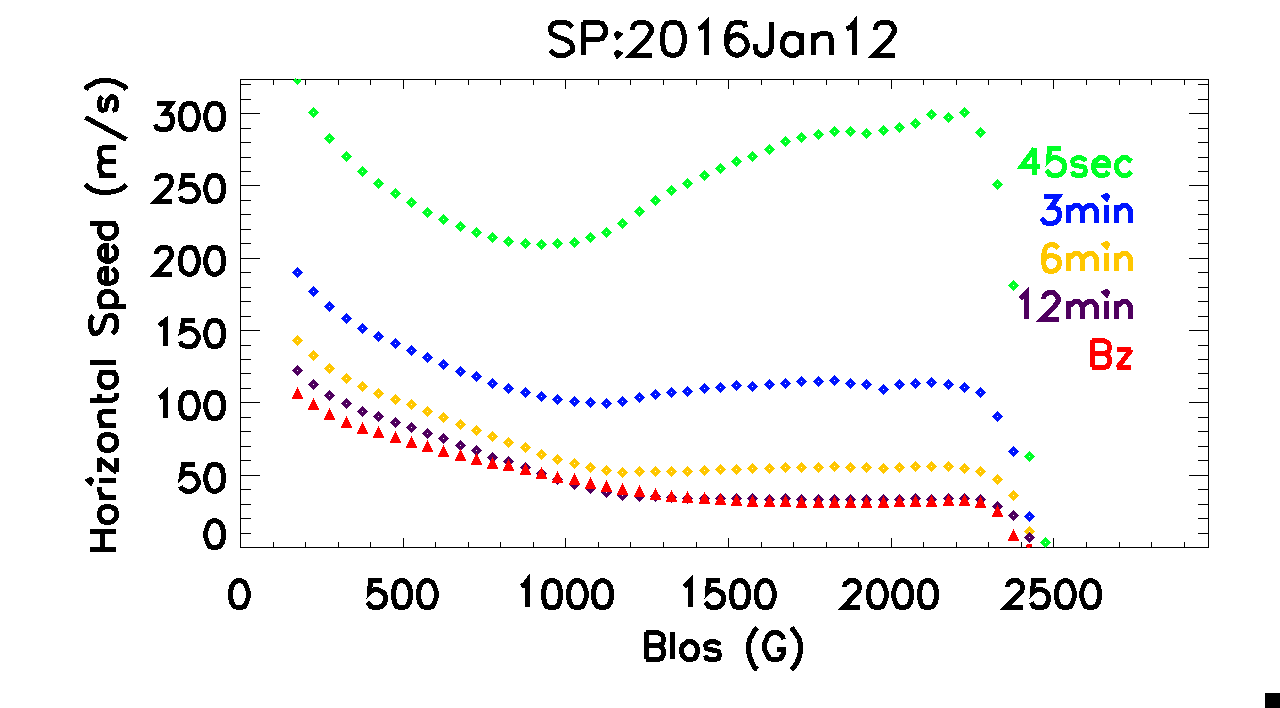}
    
    \includegraphics[trim={3cm 0.5cm 1.0cm 0},clip,scale=0.20]{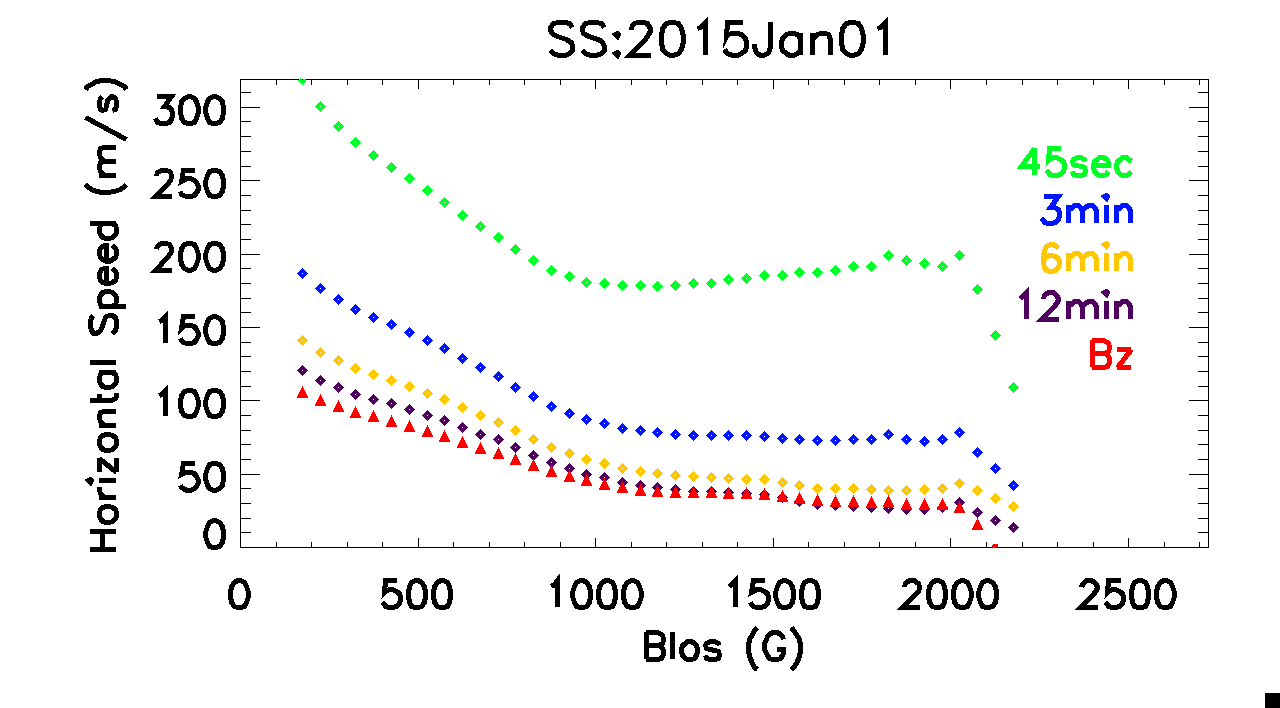}
    \includegraphics[trim={4.5cm 0.5cm 1.0cm 0},clip,scale=0.20]{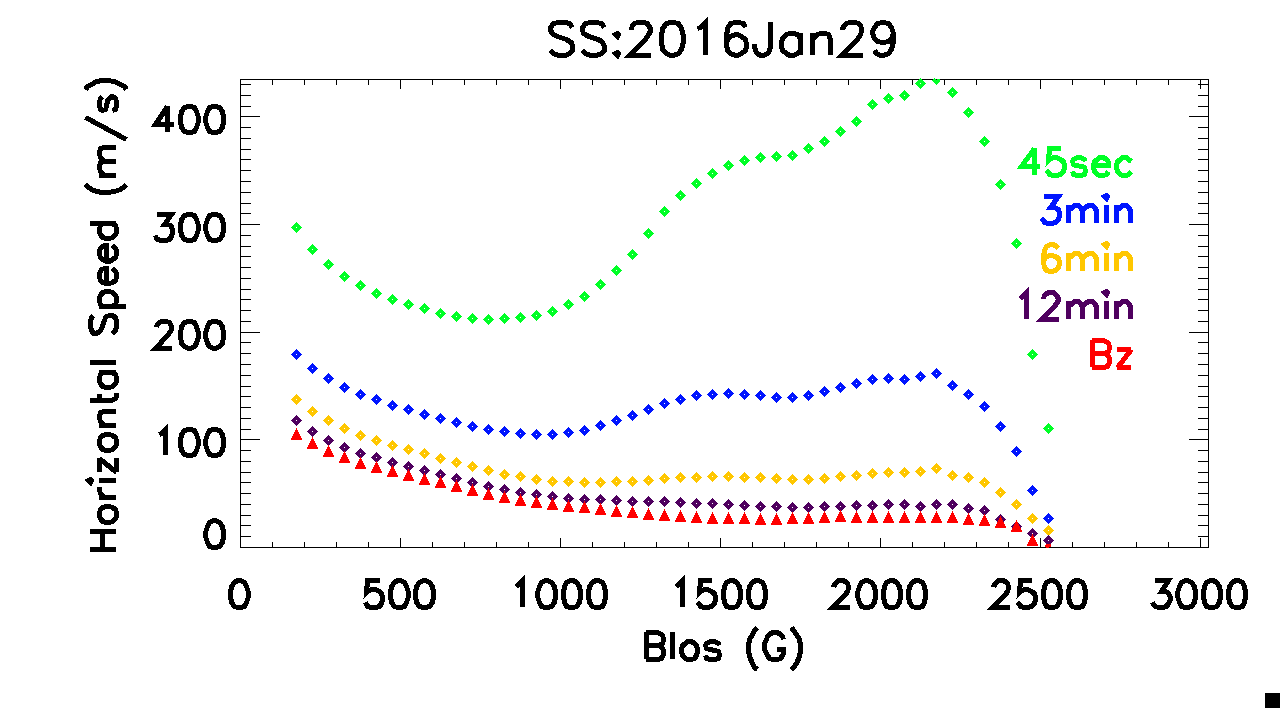}
    \caption{Horizontal speeds plotted against BLOS for SS {(left column) and SP (right column)} active regions.}
    \label{fig:horVel_BzBlos}
\end{figure}

We present the mean speeds obtained using BLOS magnetograms in Fig.\ \ref{fig:horVel_BzBlos}. FLCT was performed on BLOS magnetograms with time-steps of 45-second, 3-minute, 6-minute and 12-minutes, represented with the respective colors given in the legends (see Fig.\ \ref{fig:horVel_BzBlos}). The mean speeds are obtained the same way as from the Bz magnetograms. The mean speeds from Bz magnetograms are plotted as red triangles along with the BLOS speeds, for comparison. Although the speeds from the BLOS magnetograms are obtained with a threshold factor of 20G in FLCT, the figure only displays mean speeds for field-strengths over 150G, to maintain consistency with the Bz magnetogram curves. 
The speeds derived from BLOS magnetograms exhibit similar trends to those obtained from Bz, with speeds decreasing in general as the field strength increases, except at higher field strengths in the higher cadence curves, this is explained below. The speeds also progressively decrease in magnitude as the interval between two adjacent magnetograms increases (decreasing cadence). 
This behavior suggests that FLCT tracks shorter-, intermediate- and longer-lived advection flows with decreasing cadence. 

We performed FLCT measurements on BLOS magnetograms with 45-second and 12-minute cadences as a consistency check for each AR, because the mean speeds obtained using Bz are two to three times lesser compared to the speeds in \citet{1992ApJ...393..782T} and others. The mean speeds obtained using the 45-second data are similar to those obtained in \citeauthor{1992ApJ...393..782T}, with their image cadence (60-seconds) similar to ours. The 12-minute BLOS mean speeds are also consistent with those from Bz; the small differences of these curves are believed to be an effect of the deprojection and the smoothing of the SHARP-CEA Bz data. The 3-minute and 6-minute curves exhibit intermediate speeds suggesting that FLCT tracks features of intermediate lifetimes.
The green curves in Fig.\ \ref{fig:horVel_BzBlos} corresponding to the 45-second dataset show an increase in the speeds beyond $\sim$1000 G. This is perhaps due to contributions from umbral and penumbral p-mode oscillations, which are stronger in the 45-second dataset and are smoothed out for the 12-minute magnetograms \citep{2013SoPh..287..279L}. 
We suggest that the p-mode oscillations horizontally displace the fields, causing the tracker to follow the resulting BLOS flux displacements in the higher-cadence magnetograms. The speed-bump appears only at higher field-strengths, suggesting that at lower field strengths the advection speed of the magnetic flux from convection dominates that due to the oscillations. 
However it cannot be ruled out that p-mode oscillations drive significant coronal heating in coronal loops rooted in penumbra.

\begin{figure}[H]
    \includegraphics[trim={0cm 0.5cm 1.4cm 0},clip,scale=0.14]{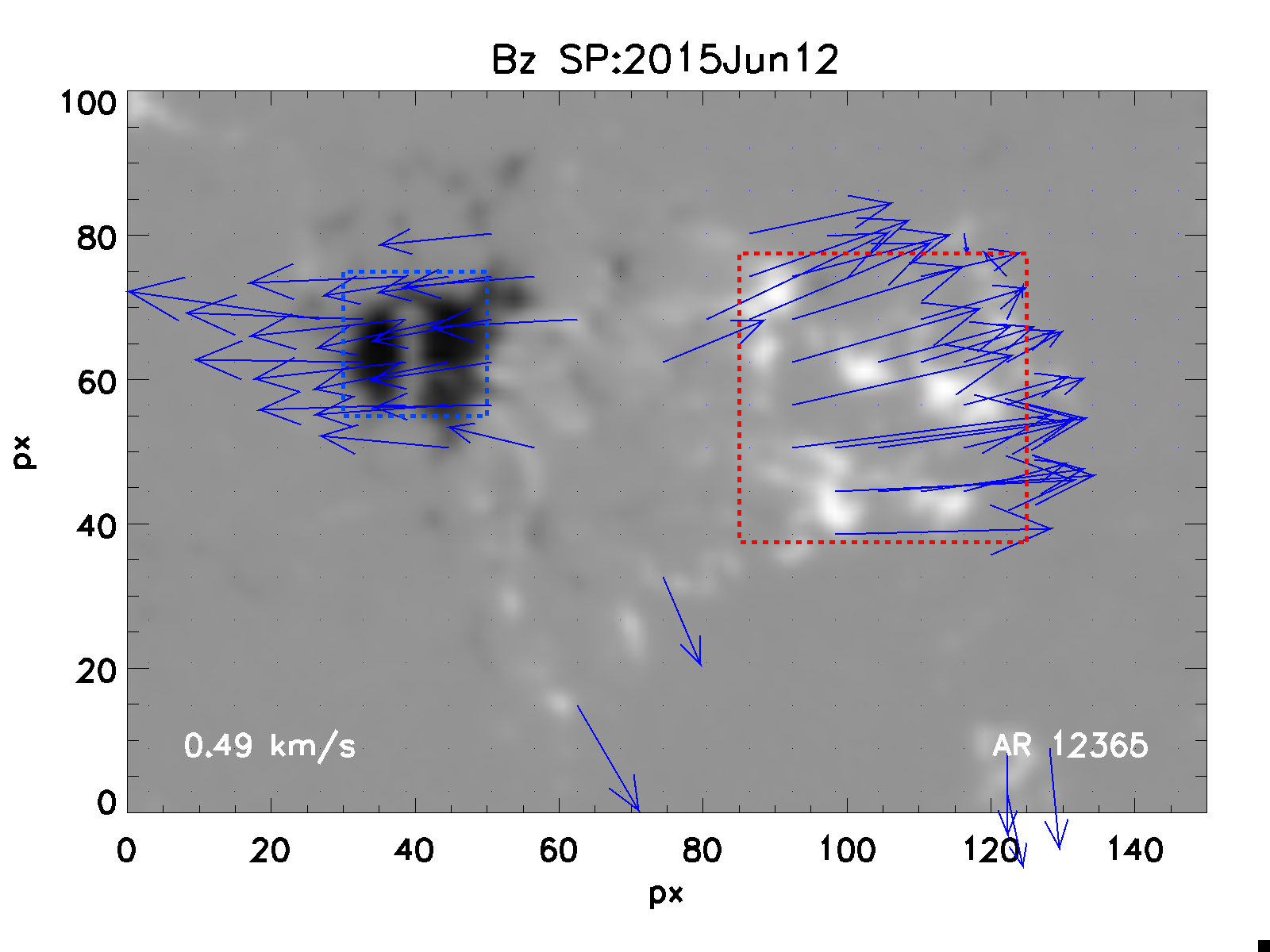}
    \includegraphics[trim={2cm 0.8cm 1.2cm 0},clip,scale=0.26]{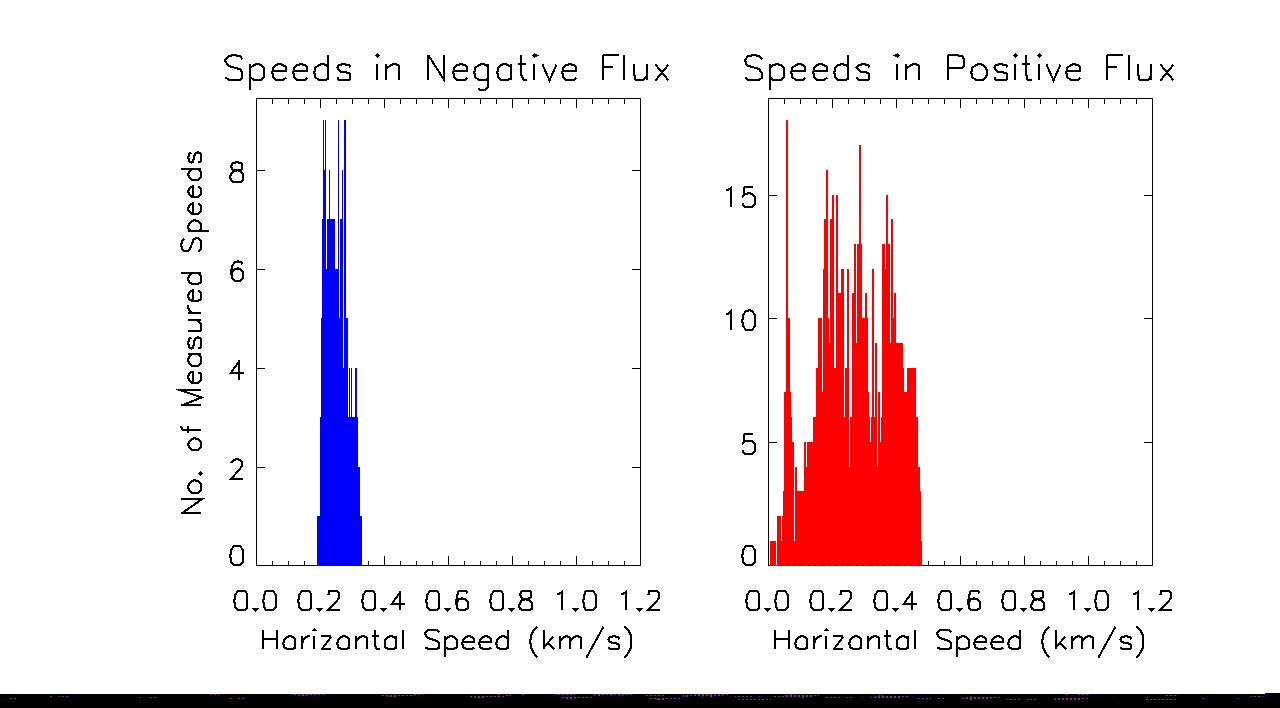}
    \caption{{Example of an emerging bipolar magnetic region in which the positive and negative flux domains each have a wide range of Bz but a narrow range of horizontal advection speed (these aspects plausibly produce the horizontal streaks in Fig.\ \ref{fig:SS_colorsAll_hist}).  
    Left:\ A close-up of the Bz magnetogram in the white box in the top right panel of Fig.\ \ref{fig:arrowsPlot}, which panel is the HMI SHARP Bz magnetogram for AR 12365 at 07:36:07 UT on 2015 June 13.  
    The blue arrows show the horizontal advection speed and direction of flux clumps in the close-up  magnetogram at that time. For the longest arrow, the speed is 0.49 km/s. In the negative domain, the Bz of the tracked flux ranges from 150G to 1159G. In the positive domain the Bz of the tracked flux ranges from 150G to 850G. 
    Right: Histograms of the horizontal advection speeds measured at that time for the negative and positive flux clumps in the blue and red dashed boxes in the magnetogram, respectively.  
    }}
    \label{fig:magnifiedImg} 
\end{figure}

\begin{figure}[H]
        \centering
         \includegraphics[trim={6.5cm 1cm 7cm 0cm},clip,scale=0.4]{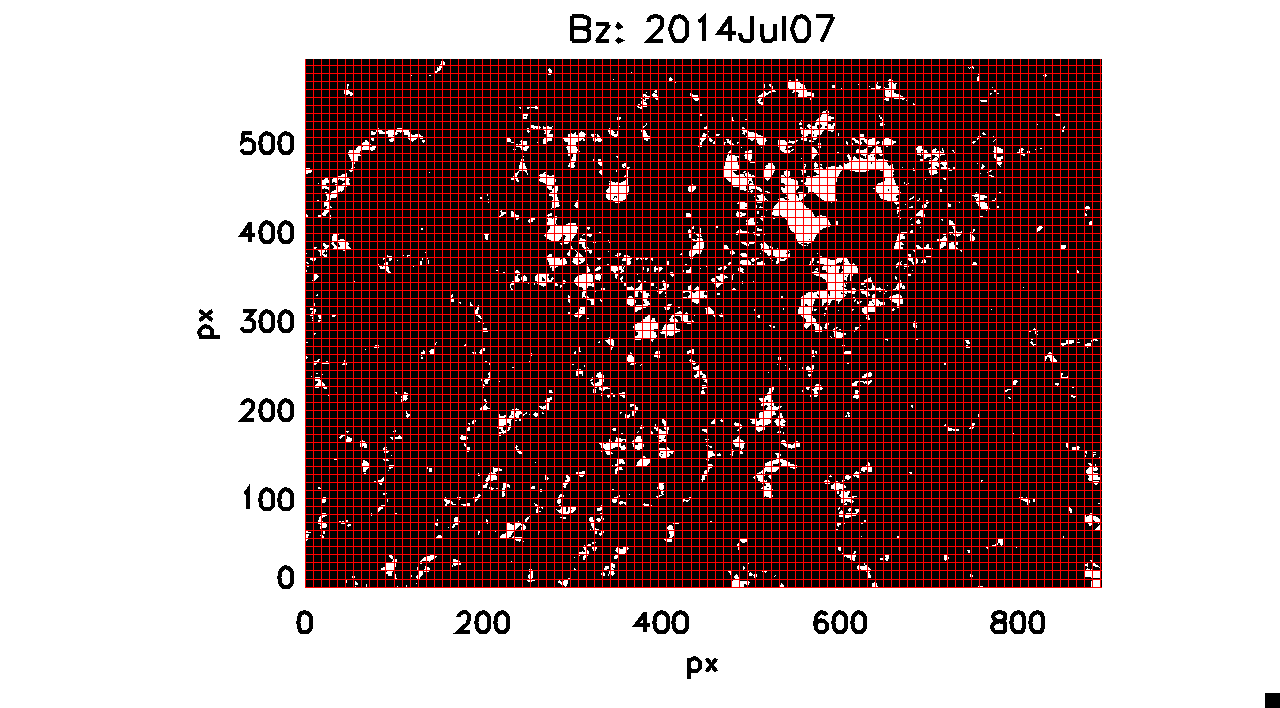}
    \caption{{The binary image for the 0.1-0.4 km/s speed range in the 2014Jul07 SS AR, with white regions covering all the pixels in that speed range in the 2-hour duration. Red boxes show the grid of 9x9 pixel blocks covering the binary image. The dots on the magnetograms in Fig.\ \ref{fig:selColLocs} are the centroids of the white pixels in each 9x9 pixel block in each binary image.  
     }}\label{fig:binaryImg}
\end{figure}

\bibliography{sample631}{}
\bibliographystyle{aasjournal}

\end{document}